\documentclass[12pt]{article}
\usepackage{amssymb,amsmath,amsthm,mathrsfs,latexsym,amsxtra,graphicx,appendix,epstopdf,feynmf,subfig,hyperref,setspace,fix-cm,color,cite,hyperref}
\numberwithin{equation}{section}
\begin{document}

\newcommand{\comment}[1]{{\bf\color{blue}[#1]}}

\newcommand{\bea}{\begin{eqnarray}}
\newcommand{\eea}{\end{eqnarray}}
\newcommand{\be}{\begin{equation}}
\newcommand{\ee}{\end{equation}}
\newcommand{\eq}[1]{(\ref{#1})}

\newcommand{\del}{\partial}
\newcommand{\delbar}{\overline{\partial}}
\newcommand{\zbar}{\overline{z}}
\newcommand{\wbar}{\overline{w}}
\newcommand{\vbar}{\overline{\varphi}}

\newcommand{\hf}{\frac{1}{2}}
\newcommand{\qrt}{\frac{1}{4}}
\newcommand{\Z}{{\mathbb Z}}
\newcommand{\R}{{\mathbb R}}
\newcommand{\C}{{\mathbb C}}
\newcommand{\A}{{\mathbb A}}
\newcommand{\N}{{\mathbb N}}
\newcommand{\bH}{{\mathbb H}}
\renewcommand{\P}{{\mathbb P}}
\newcommand{\M}{{\cal M}}
\newcommand{\Q}{{\mathbb Q}}
\newcommand{\tX}{\widetilde{X}}
\newcommand{\mO}{\Omega}
\newcommand{\mJ}{{\mathbb J}}
\def\taubar{\overline{\tau}}
\def\Tr{{\rm Tr}}
\def\qhat{\hat{q}_0}

\def\eg{{\it e.g.~}}
\def\a{\alpha} 
\def\b{\beta} 
\def\g{\gamma} 
\def\G{\Gamma}
\def\e{\epsilon}
\def\h{\eta}
\def\th{\theta} 
\def\Th{\Theta}  
\def\k{\kappa}
\def\la{\lambda} 
\def\L{\Lambda} 
\def\m{\mu}
\def\n{\nu}
\def\r{\rho} 
\def\s{\sigma} 
\def\t{\tau}
\def\f{\phi} 
\def\F{\Phi} 
\def\w{\omega}
\def\W{\Omega} 
\def\v{\varphi} 
\def\z{\zeta}

\def\lieg{{\mathfrak g}}

\def\i{{ i}}

\newcommand{\cB}{{\cal B }}
\newcommand{\cW}{{\cal W }}            
\newcommand{\cM}{{\cal M }}            
\newcommand{\cF}{{\cal F }}            
\newcommand{\cC}{{\cal C }}
\newcommand{\cL}{{\cal L }}                
\newcommand{\cO}{{\cal O }}            
\newcommand{\cH}{{\cal H }}            
\newcommand{\cA}{{\cal A }}            
\newcommand{\cG}{{\cal G }}
\newcommand{\cN}{{\cal N }}            
\newcommand{\cY}{{\cal Y }}    
\newcommand{\cD}{{\cal D }} 
\newcommand{\cV}{{\cal V }}    
\newcommand{\cJ}{{\cal J }}    
\newcommand{\E}{{\cal E }}   
\newcommand{\B}{{\cal B}} 
\renewcommand{\d}{{\partial}}

\newcommand{\wl}{{\widetilde{\lambda}}}

\def\wt{\widetilde}

\newcommand{\Gr}{\ensuremath{\mbox{Gr}}}
\newcommand{\SG}{\ensuremath{\mbox{SG}}}
\newcommand{\TN}{\ensuremath{\mbox{TN}}}
\newcommand{\CY}{\ensuremath{\mbox{CY}}}

\newcommand{\vac}{|0\rangle}

\newcommand{\ie}{{\it i.e.~}}
\newcommand{\etc}{{\it etc.~}}

\newcommand{\one}{\mathbf{1}}
\newcommand{\two}{\mathbf{2}}
\newcommand{\three}{\mathbf{3}}
\newcommand{\eo}{\epsilon_1}
\newcommand{\et}{\epsilon_2}
\newcommand{\bp}{\mathbf{+}}
\newcommand{\bm}{\mathbf{-}}

\newcommand{\wb}{{\bar w}}
\newcommand{\zb}{{\bar z}}
\newcommand{\xb}{{\bar x}}
\newcommand{\hb}{{\bar h}}
\newcommand{\qb}{{\bar q}}

\newcommand{\Vp}{V^\perp} 
\newcommand{\Vpd}{W^\perp} 

\newcommand{\rep}{R^G} 

\newcommand{\pl}[1]{\Psi^{(#1)}}
\newcommand{\pbl}[1]{\bar{\Psi}^{(#1)}}
\newcommand{\pr}[1]{\tilde{\Psi}^{(#1)}}
\newcommand{\pbr}[1]{\tilde{\bar{\Psi}}^{(#1)}}

\newcommand{\Xl}[1]{\partial X^{(#1)}}
\newcommand{\Xbl}[1]{\partial\bar{X}^{(#1)}}
\newcommand{\Xr}[1]{\partial\tilde{X}^{(#1)}}
\newcommand{\Xbr}[1]{\partial\tilde{\bar{X}}^{(#1)}}

\newcommand{\mm}[1]{\mathbf{m}^{(#1)}}
\newcommand{\ma}[1]{m^{(#1)}}

\newcommand{\gs}{|\sigma^{--}\rangle}

\newcommand{\mg}{{\ensuremath{\mathbb{M}_{24}}}}
\newcommand{\no}[1]{\!:\! #1\!\! :}

\newcommand{\Do}{D^{(1)}}
\newcommand{\Dtw}{D^{(2)}}
\newcommand{\Dth}{D^{(3)}}

\begin{titlepage}
\begin{center}

\hfill \\
\hfill \\
\vskip 0.75in

{\Large 
	\bf Lifting $\textstyle\frac14$-BPS States on K3 and Mathieu Moonshine
}\\

\vskip 0.4in

{\large Christoph A.~Keller${}^{a}$ and Ida G.~Zadeh${}^{b}$
}\\
\vskip 4mm

${}^{a}$
{\it Department of Mathematics, University of Arizona, Tucson, AZ 85721-0089, USA} \vskip 1mm
${}^{b}$
{\it International Centre for Theoretical Physics, Strada Costiera 11, 34151 Trieste, Italy} \vskip 1mm

\texttt{cakeller@math.arizona.ch,~zadeh@ictp.it}

\end{center}

\vskip 0.35in

\begin{center} {\bf ABSTRACT } \end{center}
The elliptic genus of K3 is an index for the $\textstyle\frac14$-BPS states of its $\sigma$-model. At the torus orbifold point there is an accidental degeneracy of such states. We blow up the orbifold fixed points using conformal perturbation theory, and find that this fully lifts the accidental degeneracy of the $\textstyle\frac14$-BPS states with $h=1$. At a generic point near the Kummer surface the elliptic genus thus measures not just their index,  but counts the actual number of these BPS states. We comment on the implication of this for symmetry surfing and Mathieu moonshine.
\vfill

\noindent \today

\end{titlepage}

\setcounter{tocdepth}{2}

\tableofcontents

\section{Introduction and summary}
\subsection{K3 Mathieu moonshine}
The K3 surface is of great interest in mathematics and physics \cite{Aspinwall:1996mn}. As a Calabi-Yau manifold, it is believed to lead to a 2d CFT with small $\cN=(4,4)$ supersymmetry through its non-linear $\sigma$-model. This has played an important role in many compactifications of string theory. This $\sigma$-model has also played an important role in holography, where its symmetric orbifold provides a precise instance of the AdS$_3$/CFT$_2$ correspondence \cite{Aharony:1999ti,David:2002wn}. More recently, K3 has come to prominence through its role in Mathieu moonshine. Ref. \cite{Eguchi:2010ej} observed that the decomposition of the elliptic genus of K3 into the characters of the small  $\cN=4$ algebra gives multiplicities which correspond to dimensions of representations of the Mathieu group \mg. Further evidence for this was found by checking that the characters of these representations satisfy the expected modular transformation properties \cite{Cheng:2010pq,Gaberdiel:2010ch,Gaberdiel:2010ca,Eguchi:2010fg}. Ref. \cite{Gannon:2012ck} then established the existence of a Mathieu moonshine module $V^{\mg}= \bigoplus_{n\geq 1} V^{\mg}_n$ whose graded characters have the right modular transformation properties. Even though in principle this gives an explicit construction of the moonshine module, it is not at all a natural one. The goal of understanding \mg~ moonshine is to find a natural construction, in the vein of the construction of the monstrous moonshine module as a holomorphic vertex operator algebra (VOA) \cite{MR747596,MR1172696}. An obvious approach to such a natural construction is of course to try to use the non-linear $\sigma$-model of K3.

When following this approach, several technical issues arise. The moduli space $\M$ of K3 $\sigma$-models has $\dim_\R \M =80$. More precisely, it is given by the quotient $\M = O(4)\times O(20)\backslash O(4,20)/O(4,20;\Z)$ \cite{Aspinwall:1996mn}. One issue is that for a generic point in this moduli space, we do not have an explicit construction for the $\sigma$-model. Only in places with enhanced symmetries do we know explicit descriptions: for instance at the Gepner points \cite{Gepner:1987qi,Gepner:1987vz}, where the CFT is given by a rational CFT, and for the case where K3 is a Kummer surface, \ie a torus orbifold $\mathbb T^4/\Z_2$. Away from these points, we only know the spectrum for states of high enough supersymmetry: $\textstyle\frac12$-BPS states are protected, so that their number is constant on the moduli space. Their multiplicities are in fact given by the Hodge diamond of K3. $\textstyle\frac14$-BPS states, on the other hand, are not completely protected, and their multiplicity can change. Their index however is protected. The elliptic genus counts the index, and is therefore constant on the moduli space. More precisely, when moving around on the moduli space, it is possible for two $\textstyle\frac14$-BPS states to pair up to form a non-BPS state, which then no longer contributes to the elliptic genus.  

At a `generic' point of the moduli space, we would expect that all $\textstyle\frac14$-BPS that can be lifted, will be lifted. That is, we expect the index to count the actual number of states. (This could be taken as a definition of what `generic' means). The issue with this statement is that points of moduli space where we have an explicit CFT description are not generic. Take a $\mathbb T^4/\Z_2$ orbifold, for instance. For sufficiently generic radii and $B$-fields of $\mathbb T^4$, the number of $\textstyle\frac14$-BPS states of weight $h=1$ is 102. From the elliptic genus we know that the index is 90. If we move away from the orbifold point, we would thus expect that 12 of those states get lifted. The main result of this paper is to confirm this expectation by performing perturbation theory.

\subsection{Summary of results}
Let us discuss the locus of torus orbifolds $\M_{\rm Kummer}$ and its neighborhood in the moduli space of K3 for a moment. The action of $\Z_2$, which corresponds to $\mathcal X\mapsto -\mathcal X$, where $\mathcal X$ represents collectively all the bosonic and fermionic fields of the $\sigma$-model, has 16 fixed points on $\mathbb T^4$. The torus orbifold $\mathbb T^4/\Z_2$ is thus singular as a geometry. The corresponding CFT is, however, perfectly regular. What does $\M$ look like near $\M_{\rm Kummer}$? We can of course deform the torus itself without leaving $\M_{\rm Kummer}$. From the point of view of the CFT with small $\cN=(4,4)$ supersymmetry, these deformations correspond to 16  moduli, all of which come from the untwisted sector. Alternatively, we may leave $\M_{\rm Kummer}$ by blowing up its singularities. This corresponds to turning on the moduli in the $\Z_2$ twisted sector of the CFT. There are 4 such moduli for each fixed point  --- in the $\cN=(2,2)$ language these correspond to the choice of $(c,c)$, $(c,a)$, $(a,c)$, and $(a,a)$ moduli. In total, we obtain $16+4\times16=80$ moduli, just as expected. To reach a generic point near $\M_{\rm Kummer}$, we thus start out with a torus orbifold CFT and then perturb by a twisted modulus. We want to identify which of the 102 $\textstyle\frac14$-BPS states get lifted in doing so, and which ones remain to form the space $V_1^\mg$. 

We find the following result: At the orbifold point the $\textstyle\frac14$-BPS states of dimension $h=1$ decompose into untwisted states and twisted states, 
\be
V^{\rm BPS}_1= U_1 \oplus T_1\ ,
\ee
where $\dim U_1 =6$ and $\dim T_1 =96$. We establish that under perturbation by any of the twisted moduli, all 6 states in $U_1$ and 6 states in $T_1$ get lifted. That is, their right-moving short representations combine into a long non-BPS representation, whose conformal weight (in the NS sector) is given by
\be\label{right_dimension}
\bar h(\lambda)=\frac12+\frac{\lambda^2\pi^2}{2}+O(\lambda^3)\ ,
\ee
with $\lambda$ the coupling constant. Moreover we identify precisely which of the 96 twisted states get lifted for a given modulus. If we choose the twisted modulus to be a linear combination of the 16 fixed point sectors $\beta$, that is 
\be
\cO = b^{\beta}\cO_\beta\ ,
\ee
with $b^{\beta}$ the 16-dimensional unit vector that fixes the direction in moduli space, then the 6 twisted states parallel to $b^\beta$ get lifted, and the 90 states orthogonal to that direction remain $\frac14$-BPS. In summary, this establishes that all states that can be lifted will be lifted. In the process, we also confirm that a generic point (at least near the orbifold point), the index does indeed count the number of BPS states.

This results has implications for Mathieu moonshine. Starting from the elliptic genus, the most natural guess for $V^\mg$ is to take it to be given by the space of $\textstyle\frac14$-BPS states $V^{\rm BPS}$. If we try to construct this space say at the orbifold point, we run into an immediate problem: the dimension of this space is simply too big. To obtain the right dimension, we want to restrict to states that are not lifted. Refs. \cite{Taormina:2013jza, Taormina:2013mda} make a concrete proposal for this. They pick a particular linear combination $b^\beta$ of twisted moduli, and conjecture that the correct definition for $V^\mg_1$ is the subspace of $\textstyle\frac14$-BPS states which are not lifted. In this article we construct this subspace explicitly, and check that its dimension is indeed 90. 
For the choice of modulus proposed in \cite{Taormina:2013jza, Taormina:2013mda}, this establishes that $V^\mg_1$ indeed has the expected form.

In order to establish (\ref{right_dimension}), we need to go to second order in conformal perturbation theory. The reason for this is easy to see: the conformal weight $h$ of BPS states saturates the unitarity bound. If we perturb by $\lambda$, then $h(0)$ has to be a minimum of $h(\lambda)$, which of course implies that the first order term vanishes.  This complicates the computation, since due to conformal symmetry first order perturbation theory is much simpler than higher order perturbation theory.

Nonetheless we are able to obtain exact results at second order perturbation theory thanks to supersymmetry: even though $\textstyle\frac14$-BPS states do not have enough supersymmetry to protect their conformal weight, there is enough supersymmetry so that we can use superconformal Ward identities to express their correlation functions as a total derivative. Using Stokes' theorem, we can  reduce the perturbation theory integrals over the full complex plane to contour integrals around the singularities, which in turn can be evaluated simply by computing the first few terms in the OPE of fields.

In this paper we study deformation of 2d supersymmetric CFTs, namely $\sigma$-models on K3 and rely on supersymmetry to compute lifting of the states. The approach and methodology we develop here may naturally be applied to study deformations of $\mathcal N=4$ sumersymmetric CFTs in general. One interesting example of such theories arises in the context of the AdS$_3$/CFT$_2$ correspondence. In particular, for compactification of type IIB superstring theory on K3, the dual CFT is believed to be the symmetric product orbifold of K3 $\sigma$-model. The elliptic genus of symmetric product orbifold CFTs is derived in \cite{Dijkgraaf:1996xw}. As one deforms the theory away from the orbifold point with exactly marginal operators, some $\textstyle\frac14$-BPS states acquire anomalous dimensions and leave the elliptic genus. The deformation analysis is exactly similar to what we present in this paper. The representation theory of the small $\mathcal N=4$ superconformal algebra, as well as the superconformal Ward identities, determine the lifting of the states. Of course, the degenarcy of the states grow very fast for symmetric product orbifolds and as such, the deformation analysis in terms of free field representation of the operators become cumbersome for higher modules.

Another interesting direction to explore is the deformation of families of non supersymmetric CFTs. In particular, one question to address is the lifting of the spectrum of such theories under deformations. We pursue this problem for the case of toric orbifolds $\mathbb T^d/\mathbb Z_2$ in upcoming work \cite{torusorbifold}.

This paper is organized as follows: in section~\ref{section_spectrum} and appendix~\ref{appendix:spectrum} we set up our notation and discuss the spectrum of the $\sigma$-model at the orbifold point. In section~\ref{section_2ndorder} we make some general remarks about conformal perturbation theory at second order and regularization schemes. In section~\ref{section_untwisted} we compute the lifting of the states in the untwisted sector, and in section~\ref{section_twist_lifting} we compute the lifting in the twisted sector for a specific direction. Section~\ref{s:symmetrysurfing} finally discusses the general picture for arbitrary directions in the moduli space and its application to symmetry surfing.  Appendix \ref{app_4pfs} contains technical details of the computations of the correlation functions. Appendix \ref{app_ward} outlines the derivation of the Ward identities we use.

\section{Spectrum of the torus orbifold $\mathbb T^4/\Z_2$}\label{section_spectrum}

\subsection{The small $\mathcal N=(4,4)$ superconformal algebra}
Let us begin by introducing some notation. We denote the left-moving complex fermions and complex bosons on $\mathbb{T}^4$
as 
\be\label{complexpsix}
\pl{i} \ , \quad \pbl{i}  \ , \qquad  \Xl{i}\ , \quad \Xbl{i}  \ ,\qquad i=1,2\ .
\ee
We will also sometimes denote these fields collectively by
\be\label{complexpsix_coll}
\Psi^{(I)}\ ,\quad\partial X^{(I)}\ ,\qquad I=1,2,3,4\ .
\ee
The right-moving fields are decorated by a tilde and have the same convention. The (anti-)commutation relations of the fields are given by
\be\label{Xcommu}
[\Xl{i}_m,\Xbl{j}_n]=m\delta^{ij}\delta_{m,-n}
\ee
\be\label{psianticommu}
\{\pl{i}_r,\pbl{j}_s \}=\delta^{ij}\delta_{r,-s}
\ee
with all other (anti-)commutators vanishing.
From these, and their right-moving counterparts, one constructs the small $\mathcal N=(4,4)$ superconformal algebra with central charge $c=6$ --- see appendix~\ref{app:N4} for more details. The small $\mathcal N=4$ primaries $|\phi\rangle$ (which we will just refer to as primaries) are defined in the NS sector as:
\be\label{sN4primary}
L_n|\phi\rangle=0\ ,\quad G_r|\phi\rangle=0\ ,\quad J_n|\phi\rangle=0\ ,\qquad n\in\mathbb Z_{>0}\ ,r\in\mathbb Z_{>0}+\frac12\ .
\ee
Chiral and anti-chiral primaries are denoted by $|c\rangle$ and $|a\rangle$, respectively, and, in addition the to the above conditions, satisfy
\be\label{sN4cprimary}
G^{+A}_{-\frac12}|c\rangle=0\qquad G^{-A}_{-\frac12}|a\rangle=0\ .
\ee
Here $G^{\alpha A}$ are the 4 holomorphic supercurrents of the algebra and the indices $\alpha$ and $A$ correspond to the $SU(2)$ R-symmetry and $SU(2)$ flavor symmetry, respectively, see eq. (\ref{Gs}). (Anti-)chiral primaries satisfy the unitary bound
\be\label{unitary}
h=|q|\ ,
\ee
where $h$ is the left-moving conformal dimension and $q$ is the R-charge, namely the spin of the holomorphic $\mathfrak{su}(2)$ R-symmetry algebra: $(L_0,J^3_0)=(q,|q|)$. {\footnote{In our conventions, the $U(1)$-charge $j$ of the $\mathcal N=2$ subalgebra of the small $\mathcal N=4$ superconformal algebra is the eigenvalue of the U(1)-current $J$ and is related to the $\mathfrak{su}(2)$ spin as $j=2q$.}}

At the $\Z_2$ orbifold point associated with the Kummer locus, the symmetry algebra is larger. In particular, there are 6 fermionic bilinear fields where 3 of them are the R-symmetry currents $J^{\pm,3}$ which are given in eq. (\ref{J}) and generate the $SU(2)$ R-symmetry. The other 3 fields, $\hat J^{\pm,3}$ in eq. (\ref{Jminus}), generate the $SU(2)$ flavor symmetry. The latter, however, will be lifted when perturbing away from the orbifold point.

In this paper we work with the elliptic genus defined by \cite{Gaberdiel:2016iyz} which is defined as the trace over the NS$\,\otimes\,\tilde{\rm R}$ sector of the K3 $\sigma$-model with the insertion of the fermion number operator $\tilde F$:
\be\label{ellgen}
\mathcal E^{\rm{NS}}_{\rm{K3}}(\tau,z)={\rm{tr}}_{\rm{NS\tilde{R}}}\Big(q^{L_0-\frac14}y^{J_0}\bar q^{\tilde L_0-\frac14}(-1)^{\tilde F}\Big)\ .
\ee
We study states with conformal dimensions $h_{\rm{NS}}=1$ and $\bar h_{\tilde{\rm R}}=\frac14$. As discussed in the introduction, there are 102 such states.

For the ease of computations, we will later perform a spectral flow transformation on the anti-holomorphic part and work in the NS$\,\otimes\,\widetilde{\rm{NS}}$ sector. We follow the notation of \cite{Schwimmer:1986mf} for the small $\mathcal N=4$ algebra. Under the spectral flow, the left-moving dimension $h$ and R-charge $q$ of a state transform as
\be\label{sf_i}
h\mapsto h^\prime=h+\eta q+\eta^2\frac{c}{24}\ ,\qquad q\mapsto q^\prime=q+\eta\frac{c}{12}\ ,
\ee
where spectral flow parameter $\eta$ is integral here and $\eta=1$ amounts to flowing between the NS and Ramond sectors. Similar transformations hold for the right-moving part. For the K3 $\sigma$-model $c=6$ and we have
\be\label{sf_ii}
h\mapsto h^\prime=h+\eta q+\frac{\eta^2}{4}\ ,\qquad q\mapsto q^\prime=q+\frac{\eta}{2}\ .
\ee

The small $\mathcal N=4$ algebra has two short (BPS) representations with R-symmetry spin $l=0$ and $l=\textstyle\frac12$, and one family of long representations. In the Ramond sector, the highest weigh state of the short representations are Ramond ground states and have dimension $h=\textstyle\frac14$.

The elliptic genus is an index that has contributions from $\textstyle\frac12$-BPS as well as $\textstyle\frac14$-BPS states. More precisely, it only gets contributions from right moving Ramond ground states, as seen in eq. (\ref{ellgen}). Let us for the moment consider the elliptic genus in the R$\,\otimes\,\widetilde{\rm{R}}$ sector. The $\textstyle\frac12$-BPS states are Ramond ground states for both left and right-movers, and are counted by the Hodge numbers of K3. The $\textstyle\frac14$-BPS states are Ramond ground states for the right-movers, non-BPS states for the left-movers, and have dimensions
\be
(h_R,\tilde h_R) = (\textstyle\frac14+n,\textstyle\frac14)\ ,\qquad  n>1\ .
\ee

We denote the space of such states by $V^{\rm BPS}_n$. $\rm V^{BPS}_n$ splits into states in the untwisted sector, which we will denote by $U_n$, and states in the twisted sector, which we will denote by $T_n$. Note that such a state can have two possible right-moving short representation: $\tilde l=0$, the singlet under the R-symmetry, and $\tilde l=\textstyle\frac12$, the doublet.  In total we thus have
\be\label{VUT}
V^{\rm BPS} = U^{\tilde l=0}\oplus U^{\tilde l=\frac12}\oplus T^{\tilde l=0}\oplus T^{\tilde l=\frac12}\ .
\ee
We will see below that actually $T^{\tilde l=\frac12}=0$.

We shall now consider spectral flowing to the left-moving NS sector to obtain the index (\ref{ellgen}). Using eq. (\ref{sf_ii}), we find that under spectral flow with $\eta=1$, the short representations get mapped as:
\bea\label{sf_NScps}
|h=\textstyle\frac14,q=\pm\frac12\rangle_{\rm{R}} &\to& |0,0\rangle_{\rm NS} \oplus |1,1\rangle_{\rm{NS}}\ ,\\
|h=\textstyle\frac14,q=0\rangle_{\rm{R}} &\to& |\textstyle\frac12,\textstyle\frac12\rangle_{\rm{NS}}\ .
\eea
As expected, the NS sector representations have chiral primaries as their highest weight states. The doublet gets mapped to the vacuum representation and its $J_{-1}^+$ descendant, and the singlet gets mapped to the $l=\textstyle\frac12$ chiral primary. As for the long representations of weight $n+\frac{1}{4}$ in the Ramond sector, one simply finds that they get mapped to long NS representations of weight $n$.

\subsection{Untwisted sector}\label{subsection_untwisted}
Let us now analyze in detail the states in $V^{\rm BPS}_1$. The spaces $U_n^{\tilde l}$ and $T_n^{\tilde l}$ in this subsection and the next one are defined to be in the NS$\otimes\tilde{\rm{R}}$ sector. The superscript $\tilde l$ denotes the right-moving short representation and subscript $n$ simply refers to the left-moving dimension of the highest weight state of the associated multiplet. We first study the untwisted sector and then analyze the twisted sector in the next subsection. In either case, we start from the ground states $U_0$ and $T_0$ and construct the orbifold invariant spaces $U_1$ and $T_1$.

Let us consider the untwisted sector $U^{\tilde l=0}$. Since $|\tilde l=0\rangle_{\tilde{\rm{R}}}$ is orbifold odd {\footnote{This is because we choose the ground state $|0\rangle_{\rm NS}\otimes|0\rangle_{\tilde{\rm R}}$ to be orbifold even. The Ramond ground states in the $\tilde l=0$ representation are of the form $|0\rangle_{\rm NS}\otimes\tilde\Psi^{(1)}_0|0\rangle_{\tilde{\rm R}}$ and $|0\rangle_{\rm NS}\otimes\tilde\Psi^{(2)}_0|0\rangle_{\tilde{\rm R}}$, whereas the $\tilde l=\textstyle\frac12$ representation ground states are $|0\rangle_{\rm NS}\otimes|0\rangle_{\tilde{\rm R}}$ and $|0\rangle_{\rm NS}\otimes\tilde\Psi^{(1)}_0\tilde\Psi^{(2)}_0|0\rangle_{\tilde{\rm R}}$. Thus, the $\tilde l=0$ states are orbifold odd and the $\tilde l=\textstyle\frac12$ states are orbifold even. We refer the reader to section 2 of \cite{Gaberdiel:2016iyz} for more details.\label{footevenodd}}}, there is no state in $U^{\tilde l=0}_0$.

For $U^{\tilde l=0}_{1/2}$, however, we do have the four descendants $\pl{I}_{-1/2}$ in the left-moving sector. We then spectral flow to the right-moving NS sector and obtain a total number of 16 states $\pl{I}_{-1/2}\pr{J}_{-1/2}\vac$, which are all (anti-)chiral primaries and have dimensions $(\textstyle\frac12,\textstyle\frac12)$. In the context of the small $\mathcal N=(4,4)$ algebra, they form four multiplets. The associated $G^{-A}_{-1/2}$ and $\tilde G^{-A}_{-1/2}$ descendants of the highest weight states correspond to the 16 moduli in the untwisted sector that deform the shape of the torus.

We next consider $U^{\tilde l=0}_1$. There are four states of dimension $h=1$, namely the left-moving $\Xl{I}_{-1}$ descendants. Note, however, that these are all $\mathcal N=4$ descendants of the primaries in $U^{\tilde l=0}_{1/2}$. It follows then that there are no primaries, so then we obtain
\be\label{Ul01}
U_1^{\tilde l=0}=0\ .
\ee

Next consider the $U^{\tilde l=1/2}$ sector. Since $|\tilde l=\textstyle\frac12\rangle_{\tilde{\rm{R}}}$ is orbifold even (see footnote \ref{footevenodd}), we have the two vacua in $U_0^{\tilde l=1/2}$, and no states in $U_{1/2}^{\tilde l=1/2}$. Subsequently, $U_1^{\tilde l=1/2}$ has 6 states given by the action of left-moving fermionic bilinears $\Psi^{(I)}_{-1/2}\Psi^{(J)}_{-1/2}$, see eq. (\ref{complexpsix_coll}).

Let us discuss their right-moving structure. Under spectral flow with $\eta=1$, the right-moving $\tilde l=\textstyle\frac12$ doublet decomposes into $|0,0\rangle_{\widetilde{\rm{NS}}} \oplus |1,1\rangle_{\widetilde{\rm{NS}}}$ in the NS sector, see eq. (\ref{sf_NScps}). Concentrating on $|0,0\rangle_{\widetilde{\rm{NS}}}$ for the moment, we have 6 holomorphic states
\be\label{h1hb0ut}
\Psi^{(I)}_{-1/2}\Psi^{(J)}_{-1/2}\vac_{\rm NS}\otimes\vac_{\widetilde{\rm NS}}
\ee 
with conformal dimensions $(1,0)$. Note that they correspond to the 6 currents $J^{3,\pm}$ and $\hat J^{3,\pm}$, see appendix \ref{app:N4}. However, only the latter 3 are primary fields with respect to the small $\mathcal N=4$ algebra. The former 3 are simply the R-currents, namely the $J^{3,\pm}_{-1}$ descendants of the vacuum. We thus expect $\hat J^{3,\pm}$ to be lifted, also because the enhanced symmetry at the Kummer point should disappear and only leave the actual small $\mathcal N=4$ symmetry behind. We will confirm this in section~\ref{subsection_holom}.

Similarly, for $|1,1\rangle_{\widetilde{\rm{NS}}}=\tilde{\Psi}^{(1)}_{-1/2}\tilde{\Psi}^{(2)}_{-1/2}\vac_{\widetilde{\rm NS}}$ state we have 6 non-holomorphic states
\be\label{h1hb12ut}
\Psi^{(I)}_{-1/2}\Psi^{(J)}_{-1/2}\vac_{\rm NS}\otimes\tilde{\Psi}^{(1)}_{-1/2}\tilde{\Psi}^{(2)}_{-1/2}\vac_{\widetilde{\rm NS}}\ ,
\ee 
where again the 3 states with the left-moving part $J^{3,\pm}$ are protected whereas the remaining 3 states with the left-moving part $\hat J^{3,\pm}$ are lifted away from the Kummer surface. We will analyze this in see section~\ref{subsection_nholom}.

In summary, in the $U^{\tilde l=1/2}_1$ space we have 3 $\cN=4$ primary states in (\ref{h1hb0ut}) which are of the form
\be\label{NS_NS_states1}
|1,0\rangle_{\rm{NS}}\otimes|0,0\rangle_{\tilde{\rm{NS}}}
\ee
and 3 $\cN=4$ primary states in (\ref{h1hb12ut}) of the form
\be\label{NS_NS_states2}
|1,0\rangle_{\rm{NS}}\otimes|1,1\rangle_{\tilde{\rm{NS}}}\ .
\ee
We compute their lifting in section \ref{section_untwisted}. Since they are in the same right-moving $\cN=4$ multiplet, we expect that they will be lifted by the same amount, which is indeed the case.

\subsection{Twisted sector}\label{subsection_twisted}
The toroidal orbifold $\mathbb T^4/\Z_2$ has 16 fixed points which we will denote by $\beta\in\Z_2^4$. For a given fixed point sector $\beta$, let us denote the twisted ground state by $\gs$. For the elliptic genus (\ref{ellgen}), the left-movers are in the twisted NS sector where there are fermionic zero modes. This means that the NS sector twisted ground state is degenerate. We therefore use the convention 
\be
\pbl{i}_0\gs=\pbr{i}_0\gs=0\ ,\qquad i=1,2\ .
\ee
Before projecting to orbifold even states, we note that for the left-movers we have a total of 4 ground states:
\be\label{NStwistedgrounddoublet}
|\sigma^{--}\rangle\ , \qquad \pl{1}_0\pl{2}_0\gs
\ee
forming a doublet under the R-symmetry and
\be\label{NStwistedgroundsinglet}
\pl{1}_0\gs\ , \qquad \pl{2}_0\gs
\ee
forming 2 singlets. Note that the right-moving fermions are in the twisted Ramond sector and therefore do not have zero modes. It follows that they are automatically in the $|\tilde l=0\rangle_{\tilde{\rm{R}}}$ representation. We therefore have
\be\label{Thalf}
T^{\tilde l=\frac12}=0
\ee
and we shall concentrate on $T^{\tilde l=0}$, see the comment below eq. (\ref{VUT}).

Let us now flow the right-movers to the NS sector. Before projecting to orbifold invariant states, we again get similar states as (\ref{NStwistedgrounddoublet}) and (\ref{NStwistedgroundsinglet}) for the right-movers, giving a total of $4\times 4=16$ states. The $\Z_2$ orbifold eliminates half of them. To get an even number of modes, the left and right movers need to have the same fermion parity. This means that in terms of R-symmetry representations, the surviving 8 states are arranged as
\be
T^{\tilde l=0}_{\frac12}=2\otimes 2 \oplus 4(1\otimes 1)\ .
\ee

The ground states $|\sigma^{\pm\pm}\rangle$ consist of a twist field and a spin field which impose the $\mathbb Z_2$ orbifold boundary conditions on bosons and fermions, respectively. We will discuss it in greater detail in subsection \ref{spinfields}. For the moment, we simply note that it has dimensions $(\textstyle\frac12,\textstyle\frac12)$. This implies that the $2\otimes2$ are (anti-)chiral and twisted (anti-)chiral primaries. The $4(1\otimes1)$, on the other hand, have $h=\textstyle\frac12$ and $q=0$, which means that they belong to long representations and do not contribute to the elliptic genus.

After spectral flowing to the NS-NS sector we have (anti-) chiral and twisted (anti-) chiral primaries contributing to the $2\otimes 2$. They have dimensions $(\textstyle\frac12,\textstyle\frac12)$, form doublets, and are $\textstyle\frac12$-BPS states. They do contribute to the elliptic genus, however, do not exhibit \mg-moonshine as they end up in the coefficient 20. Finally, we note that the descendants of the chiral primary under $G^{+A}_{-\frac12}$ and $\tilde G^{+A}_{-\frac12}$ give the four twisted sector moduli of the theory.

We next discuss $T^{\tilde l=0}_1$. To impose evenness, we need an even number of left-moving descendants. This gives 8 possibilities for the left movers, which we denote by
\be \label{eighthone}
T^{\tilde l=0}_{1}:\qquad| i j\rangle\equiv\sqrt{2}\Xl{i}_{-1/2}\pl{j}_0\gs\ , \qquad|\bar i j\rangle\equiv\sqrt{2}\Xbl{i}_{-1/2}\pl{j}_0\gs \ .
\ee
Two of these states are the small $\mathcal N=4$ descendants of $\gs$, namely they have the same holomorphic part as the moduli. As we will show in section \ref{section_twist_lifting}, they are of the form
\be
\textstyle\frac1{\sqrt2}(|\bar 1 1\rangle + |\bar 2 2\rangle) \ , \qquad \textstyle\frac1{\sqrt2}(|21\rangle - |12\rangle)\ .
\ee
These states are not lifted under deformation.

The remaining 6 states are primaries and are of the form
\be\label{NS_NS_twisted}
|\bar 1 2\rangle\ , \ |\bar 2 1\rangle\ , \ |11\rangle\ , \ |22\rangle\ ,
\ \textstyle\frac{1}{\sqrt{2}}(|\bar 11\rangle - |\bar 22\rangle)\  ,\
\textstyle\frac{1}{\sqrt{2}}(|21\rangle + |12\rangle)\ .
\ee
These states are lifted under deformation. We compute the lifting of these states in section \ref{section_twist_lifting} and find that the anomalous dimensions indeed match those of the untwisted sector states (\ref{NS_NS_states1}) and (\ref{NS_NS_states2}).

This completes the picture: the 6 untwisted sector and 6 twisted sector $\textstyle\frac14$-BPS states with dimensions $(1,\textstyle\frac12)$ (where the latter are along the direction of the perturbation) pair up and leave the elliptic genus away from the Kummer locus. Therefore, at generic points of the K3 moduli space, there exist 90 $\textstyle\frac14$-BPS states.

\subsection{Twist fields and spin fields}\label{spinfields}
Let us now discuss the vertex operators that correspond to the twisted sector states described in the previous subsection. $\sigma^{\pm\pm}(z,\bar z)$ are products of two operators: $\sigma_b(z,\bar z)$, which imposes the twist conditions for the bosons, and $\sigma^{\pm\pm}_f(z,\bar z)$ which imposes the twist conditions for the fermions.

The bosonic part is the usual twist field. For $c$ bosons it has dimensions
\be\label{hsigmab}
h_{\sigma_b}= \Big(\frac{c}{16},\frac{c}{16}\Big)\ .
\ee
Bosonic correlation functions have branch cuts between twist fields. One way to evaluate them is by mapping the problem to the covering surface \cite{Dixon:1986qv}. We summarize the basic results for 4-point functions of twist fields in appendix~\ref{app:bostwist}. Note that even though our $\Z_2$ orbifold is not a symmetric orbifold, the technology developed in, \eg \cite{Lunin:2000yv,Lunin:2001pw,Calabrese:2009qy,Pakman:2009zz,Pakman:2009ab} carries over to a great extent.

The fermionic part is often split up into a twist part and a spin part. For our purposes, it is convenient to follow \cite{Dixon:1986qv} and keep those two parts together. For fermions it is unnecessary to use the covering space method. Instead, the necessary branch cuts can be introduced by the bosonisation of the fermions. We define
\bea\label{bsnsn_2compferms}
&&\Psi^1=e^{iH^1}\ ,\qquad\bar\Psi^1=e^{-iH^1}\ ,\\
&&\Psi^2=e^{iH^2}\ ,\qquad\bar\Psi^2=e^{-iH^2}\ ,\nonumber
\eea
where $H^1$ and $H^2$ are real bosonic fields. We split up $\sigma_f^{\pm\pm}$ into left and right-moving parts. The bosonised spin fields are of the form
\be\label{bsnstn_spinfield}
\sigma_f^\pm=e^{\pm\frac i2(H^1+H^2)}\ ,\qquad\tilde \sigma_f^\pm=e^{\pm\frac i2(\tilde H^1+\tilde H^2)}\ ,
\ee
with conformal dimensions 
\be\label{hsigmaf}
h_{\sigma^\pm_f} =\bar h_{\tilde\sigma^\pm_f} = \frac{1}{4} \ .
\ee
Note that zero modes $\pl{i}_0$ simply act as multiplication by $e^{iH^i}$. Correlation functions of the $\sigma_f$ are then computed using Wick contraction. We describe this process in appendix~\ref{app:fermtwist}.

Combining eqs. (\ref{hsigmab}) for 4 bosons and (\ref{hsigmaf}) we find that 
\be\label{hsigmapm}
(h_{\sigma^{\pm\pm}},\bar h_{\sigma^{\pm\pm}})=\Big(\frac12,\frac12\Big)
\ee
as expected.

\subsection{The moduli}
Let us finally construct the moduli by acting with $G^{\alpha A}_{-\frac12}$ and $\tilde G^{\alpha A}_{-\frac12}$ descendants on the chiral primaries. We have $G^{-A}_{-\frac12}\gs=\tilde G^{-A}_{-\frac12}\gs=0$. Let us first consider the holomorphic part. Acting with the $G^{+A}_{-\frac12}$ yields two states of dimension (1,$\textstyle\frac12$):
\bea
&&O_1 = G^{+1}_{-\frac12}|\sigma^-\rangle =
(\Xbl{1}_{-\frac12}\pl{1}_0 + \Xbl{2}_{-\frac12}\pl{2}_0)|\sigma^-\rangle\ ,\label{Phisexplicit} \\
&&O_2 = G^{+2}_{-\frac12}|\sigma^-\rangle = (\Xl{2}_{-\frac12}\pl{1}_0 - \Xl{1}_{-\frac12}\pl{2}_0)|\sigma^-\rangle\ .\label{Phisexplicit_ii}
\eea
They are singlets under the left-moving R-symmetry and preserve the small $\mathcal N=4$ supersymmetry. We note that in the small $\mathcal N=4$ algebra $G^{+A}_{-1/2}\sigma^-=- G^{-A}_{-1/2}\sigma^+$. Therefore, out the 4 possible left-moving combinations $G^{\mp A}_{-1/2}\sigma^\pm$, there exist only 2 independent terms in the left-moving sector, namely eqs. (\ref{Phisexplicit}) and (\ref{Phisexplicit_ii}).

Let us also note that the states (\ref{Phisexplicit}) are not singlets under the flavor $SU(2)$ symmetry generated by $\hat J$ (\ref{Jminus}). This is expected, as we do not expect that the flavor $\hat J$ currents are conserved away from the orbifold point.
 
We perform the same procedure for the anti-holomorphic part and again obtain two states. In total, we thus get a four dimensional space of the twisted sector moduli, namely $(c,c),(a,c),(c,a),(a,a)$ in the $\mathcal N=2$ convention, at each fixed point. In total, we have $4\times16=64$ twisted sector moduli for the K3 $\sigma$-model. The actual moduli are then the hermitian combinations of the complex moduli.

For concreteness we pick the modulus to be
\be\label{ccmodulus}
O = G^{+1}_{-\frac12}\tilde G^{+1}_{-\frac12}\gs =(\Xbl{1}_{-\frac12}\pl{1}_0 + \Xbl{2}_{-\frac12}\pl{2}_0)(\Xbr{1}_{-\frac12}\pr{1}_0 + \Xbr{2}_{-\frac12}\pr{2}_0)\gs\ .
\ee
Its hermitian conjugate then reads
\be\label{ccmodulus_hc_i}
O^\dagger = G^{-2}_{-\frac12}\tilde G^{-2}_{-\frac12}|\sigma^{++}\rangle =(\Xl{1}_{-\frac12}\pbl{1}_0 + \Xl{2}_{-\frac12}\pbl{2}_0)
(\Xr{1}_{-\frac12}\pbr{1}_0 + \Xr{2}_{-\frac12}\pbr{2}_0)|\sigma^{++}\rangle\ .
\ee
Using $|\sigma^{++}\rangle=\pl{1}_0\pl{2}_0\pr{1}_0\pr{2}_0\gs$ and anti-commuting the fermions to the right, we find
\be\label{ccmodulus_hc}
O^\dagger = 
(\Xl{2}_{-1/2}\pl{1}_0 - \Xl{1}_{-1/2}\pl{2}_0)
(\Xr{2}_{-1/2}\pr{1}_0 - \Xr{1}_{-1/2}\pr{2}_0)\gs\ ,
\ee
which is indeed equal to eq. (\ref{Phisexplicit_ii}) and its right-moving counterpart, as expected (see the comment below that equation). The hermitian linear combination of the two moduli are:
\be\label{ccmodulus_real}
\cO\equiv\textstyle\frac{1}{\sqrt{2}}(O+O^\dagger)\ , \qquad \widehat{\cO}\equiv\frac{i}{\sqrt{2}}(O-O^\dagger)\ .
\ee

\section{Perturbation theory and regularization}\label{section_2ndorder}
Let us now briefly review conformal perturbation theory at second order. We are interested in the shift of conformal weights of the fields, namely their anomalous dimensions. To this end, we shall compute the perturbed two point function of a field $\varphi$:
\be\label{original2pt}
\langle \varphi^\dagger(z_1)\varphi(z_4)\rangle
= \langle \varphi^\dagger(z_1)\;\varphi(z_4)\;e^{\lambda\int d^2z \Phi(z)}\rangle\ ,
\ee
where $\Phi$ is an exactly marginal field, \ie a modulus {\footnote{We use $\Phi$ to denote the moduli of a CFT in general. The moduli of the K3 $\sigma$-model in the twisted sector of the $\mathbb Z_2$ are denoted as $O$ and $O^\dagger$, see eq. (\ref{ccmodulus}).}}, of the theory and $\varphi$ has dimensions $(h_\varphi,\bar h_\varphi)$.
The second order term in perturbation theory is thus given by 
\be\label{2ndorderint}
\frac{\lambda^2}{2} \int d^2 z_2 d^2 z_3 \langle \varphi^\dagger(z_1)\Phi(z_2)\Phi(z_3)\varphi(z_4)\rangle\ .
\ee

As usual, this integral is divergent and needs to be regularized, but let us ignore this issue for a moment (we will get back to it soon). Just as in the first order perturbation theory, we use global conformal symmetry to simplify the integral. We use the M\"obius transformation
\be
z \mapsto f(z):= \frac{(z-z_4)(z_2-z_1)}{(z-z_1)(z_2-z_4)}
\ee
to rewrite the expressions in terms of the cross-ratio $x\equiv f(z_3)$, which we use to replace $z_3$. The integration measure changes as
\be
d^2 z_2 d^2 z_3 = d^2 z_2 d^2 x 
\left| \frac{\partial(z_2,z_3)}{\partial(z_2,x)}\right|^2 
=d^2 z_2 d^2 x \left|\frac{(z_1-z_3)^2 (z_2-z_4)}{(z_1-z_2) (z_1-z_4)} \right|^2
\ee
and the integral (\ref{2ndorderint}) in total reads
\be
\frac{\lambda^2}{2}\int d^2 z_2\,(z_1-z_4)^{-2h_\varphi}(\bar z_1-\bar z_4)^{-2\bar h_\varphi}\left|\frac{z_1-z_4}{(z_1-z_2) (z_2-z_4)} \right|^2
\int d^2 x \,\langle \varphi^\dagger(\infty)\Phi(1)\Phi(x)\varphi(0)\rangle\ .
\ee

Naively, the $x$ integral is independent of $z_2$. If this were the case, we could simply evaluate the $z_2$ integral. This integral is divergent, so that we now need to regularize it. The simplest way to do that is to cut $\epsilon$-discs around $z_1$ and $z_4$. This prescription yields
\be\label{2ndorder}
\pi\lambda^2 \log\left(\frac{|z_1-z_4|^2}{\epsilon^2}\right) (z_1-z_4)^{-2h_\varphi}(\bar z_1-\bar z_4)^{-2\bar h_\varphi}
\int d^2 x \langle \varphi^\dagger(\infty)\Phi(1)\Phi(x)\varphi(0)\rangle\ .
\ee
The numerical coefficient of $\log |z_1-z_4|$ gives the shift of the conformal dimension of $\varphi$ \cite{Dijkgraaf:1987jt,Cardy:1986ie,Eberle:2001jq,Friedan:2012hi}. This implies that the shift is given by the $x$ integral of the 4-point function. This is however problematic: the $x$ integral is itself divergent and needs to be regularized as well. One would expect that a change of regularization scheme could easily change the constant part of the integral, which would imply that the shift in conformal dimension is scheme-dependent.

This, however, turns out not to be the case. The point is that we need to be more careful about the $x$ integral. Our original regularization scheme in fact \emph{does} introduce a $z_2$ dependence for the $x$ integral. More precisely, we need to regularize the integrals in (\ref{2ndorderint}) already. We regularize the $z_3$ integral by cutting out $\epsilon$-discs around $z_1,z_2,z_4$. Let us concentrate on the disc around $z_4$ first. 
After the coordinate transformation, this will cut out a disc in the $x$ integral around $x=0$. The cross-ratio is of the form
\be
x= \frac{(z_3-z_4)(z_2-z_1)}{(z_3-z_1)(z_2-z_4)}
\ee
and the radius of the disc to leading order in $\epsilon$ reads {\footnote{Considering the leading order in $\epsilon$ is enough here because in practice the singularities of the 4-point function will not be of a high order.}}
\be
|x|> \epsilon\,\frac{|z_2-z_1|}{|z_4-z_1||z_2-z_4|}=: \epsilon'\ .
\ee
The point is now that $\epsilon'$ depends on $z_2$, so that the $x$ integral indeed depends on $z_2$ if there are divergences.

Let us therefore assume that the OPE of $\Phi$ and $\varphi$ contains a relevant  field $\phi$ of dimension $(h,\bar h)$,
\be\label{OPE1}
\Phi(x)\varphi(0)\sim \frac{1}{x^{1+h_\varphi-h}\xb^{1+\hb_\varphi-\hb}}\phi(0)\ .
\ee
The integral around $0$ vanishes unless $\varphi$ and $\phi$ have the same spin, in which case it leads to the divergence
\be
\sim \epsilon'^{-\Delta_\varphi+\Delta}\ ,
\ee
where $\Delta_\varphi=h_\varphi+\hb_\varphi$ and $\Delta=h+\hb$. This leads to a $z_2$ integral of the form
\be
\sim \epsilon^{-\Delta_\varphi+\Delta} 
|z_4-z_1|^{\Delta_\varphi-\Delta+2} \int d^2z_2 
\frac{|z_1-z_2|^{-\Delta_\varphi+\Delta-2}}{ |z_2-z_4|^{-\Delta_\varphi+\Delta+2}}
\sim \frac{\epsilon^{-2\Delta_\varphi +2\Delta}}{|z_1-z_4|^{-2\Delta_\varphi +2\Delta}}+ \epsilon^{0}\ .
\ee
We see that as long as $\Delta\neq\Delta_\varphi$, the regularization of the $x$ integral does not give a contribution to the $\log |z_1-z_4|$ term. This imposes the condition that there are no fields $\phi$ with $\Delta=\Delta_\varphi$ in the OPE $\Phi\;\varphi$. Together with the condition that the spins of $\phi$ and $\varphi$ are equal, we find the stronger constraint that for
\be\label{condition}
h\ne h_\varphi\qquad{\rm or}\qquad\bar h\ne\bar h_\varphi
\ee
the $x$ integral yields no logarithmic terms.

This condition is satisfied for the states we study: for the holomorphic states (\ref{NS_NS_states1}) in the untwisted sector, there is no such OPE simply because $\phi$ will be in the twisted sector and hence, non-holomorphic, which then yields $\bar h\ne\bar h_\varphi$. For the non-holomorphic untwisted sector states (\ref{NS_NS_states2}), $\varphi$ is marginal and the condition requires that there should not be any marginal fields in the OPE. This is indeed the case because the states $\varphi$ (\ref{NS_NS_states2}) are descendants of the holomorphic states (\ref{NS_NS_states1}) by the action of the anti-holomorphic R-current mode $\tilde J^+_{-1}$ and as such, the operator $\phi$ on the RHS of OPE (\ref{OPE1}) will necessarily have $\bar h>1$ in order for the R-charge to be conserved.{\footnote{We thank the referee for pointing this out.} 
Finally, for the twisted sector states (\ref{NS_NS_twisted}) with $(h_\varphi,\bar h_\varphi)=(1,\textstyle\frac12)$, the condition (\ref{condition}) is satisfied as we will show in section \ref{section_twist_lifting}.

More generally, the condition (\ref{condition}) is simply that the OPE (\ref{OPE1}) does not contain terms $x^{-1}\xb^{-1}$ --- that is, the $x$ integral should not lead to logarithmic divergences. A similar argument leads to the same conclusion for the divergences at $x=1$ and $x=\infty$.

The conclusion is thus the following: the proper regularization scheme has to be defined on the level of the $z_2$, $z_3$ integral: that is, we cut out $\epsilon$-discs for $z_2,z_3$. The requirement to have a local regularization scheme is that these discs are independent of the other variables. Once we transform to the $x$ integral, the discs for the $x$ variable depend on the $z_i$ variables. This is crucial: we could have tried to introduce an ad hoc regularization scheme for the $x$ integral by cutting out $\epsilon$-discs for $x$. This scheme, however, would not have been local since in the original correlator the discs would have depended on the other insertions. However, by dimensional analysis, the divergent part of the $x$ integral --- \ie for terms containing negative powers of $\epsilon$ --- will always come with some powers of $(z_2-z_i)$, which will affect the $z_2$ integrals, thus not producing a $\log$ term.

It follows that only the finite part of the $x$ integral contributes to the shift in the conformal dimension of $\varphi$. This means that we can actually just use a standard regularization with $\epsilon$-discs around $0,1,\infty$, giving
\be
H(\epsilon)=\int_{D_\epsilon} d^2 x \langle \varphi(\infty)\Phi(1)\Phi(x)\varphi(0)\rangle
= \sum_n H_n \epsilon^n\ .
\ee
Inserting this in eq. (\ref{2ndorder}), the shift in the conformal dimension is found to be
\be\label{2ndordertotal}
\gamma\equiv\delta h_\varphi=\delta\bar h_\varphi=-\pi\lambda^2 \log(|z_1-z_4|^2) (z_1-z_4)^{-2h_\varphi}(\bar z_1-\bar z_4)^{-2\bar h_\varphi} H_{0}\ ,
\ee
where $H_{0}$ is the $\epsilon^0$ term in the expansion of the $x$ integral. We will compute $H_0$ in several different ways. Often, the only divergent contributions come from the vacuum, so that we can simply replace the 4-point function by its connected pieces. In other cases the integrand will turn out to be a total derivative, so then we simply evaluate the resulting contour integral and extract the $H_0$ term without worrying about convergence.

\section{Lifting the untwisted sector}\label{section_untwisted}
\subsection{Holomorphic states}\label{subsection_holom}
Let us now compute the lifting of the 6 states in the untwisted sector introduced in (\ref{NS_NS_states1}) and (\ref{NS_NS_states2}). In this subsection we concentrate on the 3 holomorphic states (\ref{NS_NS_states1}) which have  dimension $(1,0)$. We use two different methods to compute the lifting: $i)$ we use the general formula derived in \cite{Gaberdiel:2015uca} for the lifting of higher-spin currents, see their eq. (5.7). $ii)$ We compute the lifting from first principles following the regularization scheme we introduced in section \ref{section_2ndorder}. We find agreement between the two methods, as expected.

\subsubsection{Spin-1 currents}\label{subsubsection_spin1}
The lifting matrix $\gamma$ for the anomalous dimension of higher-spin fields $W^{(s)k}$, with dimensions $(s,0)$ and spin $s$, under deformation by a twisted sector modulus $O$ at second order is given by the formula \cite{Gaberdiel:2015uca}:
\be\label{anomdim}
\gamma^{k\ell}=\lambda^2\pi^2\sum_{m=1-s}^{s~\rm{mod}~1}(-1)^{\lceil s\rceil-1-\lfloor m\rfloor}{2s-2\choose s-m-1}\langle O|W^{(s)k}_{-m}\;W^{(s)\ell}_{m}|O\rangle\ .
\ee
The states we consider have $s=1$ and are bilinears in fermions, see eqs. (\ref{h1hb0ut}). In particular, we have $W_{m}|O\rangle=0$ for $m>0$. We thus just need to compute the inner product matrix of states of the form $W_0 |O\rangle$.

As a warm-up, let us first consider the 3 states in (\ref{h1hb0ut}) that are not primary fields. They are $\cN=4$ descendants, and correspond to the 3 R-currents $J^{3,\pm}$ --- see eqs. (\ref{J}). We note that the normal ordering constant for $J^3$ is $-1$, namely $J^3_0=\textstyle\frac12(\pl{1}_0\pbl{1}_0+\pl{1}_0\pbl{1}_0-1)$. We know that these currents are protected by supersymmetry across the moduli space, and should therefore not be lifted. Indeed, using formula (\ref{anomdim}), we immediately find
\be\label{Rcurrents}
\gamma^{k\ell}=||J^{\pm,3}_0 |O\rangle||^2=0\ ,
\ee
which vanishes simply by virtue of the fact that $|O\rangle$ is a singlet under the R-symmetry.

The remaining three states in (\ref{h1hb0ut}) are $\mathcal N=4$-primary fields. They correspond to the $\hat J^{\pm,3}$ that are given by (\ref{Jminus}). We expect them to be lifted: these are the currents associated with a flavor $SU(2)$ symmetry at the orbifold point which will not survive at a generic point where we do not expect any symmetries beyond small $\cN=(4,4)$. That is, the enhanced symmetry of the Kummer surface should disappear and only leave the small $\mathcal N=(4,4)$ symmetry behind.

The states whose inner product matrix we need to compute are
\bea\label{Jhatstates}
\pbl{2}_0\pl{1}_0 |O\rangle &=& \Xbl{2}_{-1/2}\pl{1}_0\tilde G^{+1}_{-1/2} \gs\ ,\\
\pbl{1}_0\pl{2}_0 |O\rangle &=& \Xbl{1}_{-1/2}\pl{2}_0\tilde G^{+1}_{-1/2} \gs\ ,\\
\frac{1}{{2}}\left(-\pl{1}_0\pbl{1}_0 +\pl{2}_0\pbl{2}_0\right)|O\rangle &=& 
\frac{1}{{2}}\left(\Xbl{1}_{-1/2}\pl{1}_0- \Xbl{2}_{-1/2}\pl{2}_0\right)\tilde G^{+1}_{-1/2} \gs\ .\quad
\eea
We shall then insert them in eq. (\ref{anomdim}) and compute the $3\times3$ matrix $\gamma^{k\ell}$. For the current $\hat J^3$, the normal ordering constant is 0: $\hat J^3_0=\textstyle\frac12(-\pl{1}_0\pbl{1}_0+\pl{2}_0\pbl{2}_0)$. We immediately find that this matrix is diagonal. Moreover, all the diagonal entries are the same:
\be\label{gammas_ut}
\gamma^{k\ell}= \frac{\lambda^2\pi^2}{2}\delta^{k\ell}\ .
\ee
In summary, as expected, the three primary fields $\hat J^{\pm,3}$ get lifted ---their weight is shifted by a positive contribution--- and the three R-currents $J^{\pm,3}$ remain unlifted. This agrees with the fact that $|O\rangle$ is a singlet of the R-symmetry, and is thus annihilated by all $J^{\pm,3}_0$, 
but it is not a singlet of the $\hat J^{\pm,3}$, which means that those matrix elements are non-vanishing, and the currents are lifted.

\subsubsection{Regularization of the second order integrals}\label{subsubsection_hol}
Let us now compute the lifting of the holomorphic fields from scratch, using the regularization scheme of section \ref{section_2ndorder}. This will serve as a warm-up for more complicated computations later on, and we will of course recover the same result as above. We start out with 
\be
\int d^2 z_2 d^2 z_3 \langle \varphi^k(z_1)O^\dagger(z_2)O(z_3)\varphi^\ell(z_4)\rangle\ ,
\ee
where $\phi^k$ and $\phi^\ell$ have the same left and right-moving dimensions $(h_\phi,\bar h_\phi)$. Using eq. (\ref{2ndorder}), we need to evaluate the regularized integral
\be\label{2norderint}
\int d^2 x\,\mathcal G^{k\ell}(x)\equiv\int d^2 x  \langle \varphi^k(\infty)O^\dagger(1)O(x)\varphi^\ell(0)\rangle\ .
\ee
For convenience, we permute the positions of the fields to
\be\label{changeG}
\mathcal G^{k\ell}(x)=\langle\varphi(\infty)O^\dagger(1)O(x)\varphi(0)\rangle= (x-1)^{2h_\varphi} (\bar x-1)^{2\bar h_\varphi} |x-1|^{-4}
\langle O|\varphi^k(1)\varphi^\ell(x)|O\rangle\ .
\ee
Note that we are using the transformation
\be
z \mapsto \frac{-1+x}{(-1+z)^2}
\ee
and the definition
\be
\langle\varphi^k(\infty)O^\dagger(1)O(x)\varphi^\ell(0)\rangle
= \lim_{\epsilon_1,\epsilon_2\to 0}\epsilon_1^{-2h_\varphi -2\bar h_{\varphi}}\langle \varphi^k(\epsilon_1^{-1})O^\dagger(1+\epsilon_2 (1-x))O(x)\varphi^\ell(0)\rangle\ .
\ee

Let us now compute the 4-point function 
$
\langle O|\varphi^k(1)\varphi^\ell(x)|O \rangle\ 
$
in (\ref{changeG})
where the states $\varphi$ are given in eq. (\ref{h1hb0ut}). We start out with the states associated with $\hat J$, namely eq. (\ref{Jminus}). For concreteness, we consider $\varphi = \pbl{2}_{-\frac12}\pl{1}_{-\frac12}\vac$, whose modes are given by
\be
\varphi_m := V_m\Big(\pbl{2}_{-\frac12}\pl{1}_{-\frac12}\vac\Big)\ .
\ee
The 4-point function then reads
\be
\langle O|\varphi(1)\varphi(x)|O \rangle
= \sum_{m=0}^\infty ||\varphi_{-m} |O\rangle||^2 x^{m-1}\ .
\ee
A straightforward computation gives 
\be\label{lifting_Jhatplus}
||\varphi_{-m} |O\rangle||^2= \left\{\begin{array}{cc} m &: m \geq 1\\ 1/2 &: 0 \end{array}\ , \right.
\ee
which yields
\be
\langle O|\varphi(1)\varphi(x)| O\rangle
= \sum_{m=0}^\infty ||\varphi_{-m} |O \rangle||^2 x^{m-1}=\frac{1}{(1-x)^2} + \frac{1}{2x}\ .
\ee
From (\ref{changeG}) we get
\be
\mathcal G(x)= \left(  \frac{1}{(1-x)^2} + \frac{1}{2x}\right)\frac{1}{(1-\bar x)^2}\ .
\ee
The resulting integrand clearly diverges at $x=1$. This is due to the presence of the vacuum in the OPE of $O$ with itself. We can regularize this by subtracting the disconnected part, which consists only of one term: $\varphi$ only contracts with $\varphi$ and not with $O$, since the latter has an odd number of fermionic generators. This gives
\be
\mathcal G_{\rm connected}(x)= \mathcal G(x)- \frac{1}{|1-x|^4} = \frac{1}{2x(1-\bar x)^2}\ .
\ee
Alternatively, we could also simply evaluate the integral and obtain a contribution $\epsilon^{-2}$ from the singularity at $x=1$. By the remarks in section~\ref{section_2ndorder}, we can simply neglect that contribution since it does not contribute to the constant part $H_0$.

Note that the antiholomorphic part of $\mathcal G_{\rm connected}(x)$ is a total derivative of the function $(1-\xb)^{-1}$. This is no coincidence and follows from supersymmetry. We will use this fact when computing the lifting of the twisted sector states in section \ref{section_twist_lifting}. This allows us to apply Stokes' theorem:
\be\label{stokes}
\int_{\partial U} F dz + G d\zb = \int_U \left(\partial_z G- \partial_\zb F\right) dz\wedge d\zb
\ee
with the complex integration measure 
\be
dx\wedge dy = \frac{i}{2} dz \wedge d\bar z\ .
\ee
This reduces the area integral to a contour integral around the points $x=0,1,\infty$:{\footnote{There is an additional minus sign in the contour integral which comes from the fact that we put clockwise contours around $0,1$, and $\infty$}}
\be\label{utpi2}
\frac{i}{2}\int dx\wedge d\bar x \frac{1}{2x(1-\bar x)^2}
= \frac{i}{4}\oint_{0,1,\infty} dx \frac{1}{x(1-\bar x)} =-\frac{\pi}{2}\ .
\ee
Of the three contours, only the integral around $x=0$ contributes. From (\ref{2ndordertotal}) we therefore get  in total
\be
\gamma=\frac{\lambda^2\pi^2}{2}\ .
\ee
Similar analysis shows that the same result holds for the other two states $\hat J^-$ and $\hat J^3$.

For the R-currents (\ref{J}), the situation is different: instead of (\ref{lifting_Jhatplus}) we now get
\be
||\varphi_{-m} |O\rangle||^2= \left\{\begin{array}{cc} m &: m \geq 1\\ 0 &: 0 \end{array}\ , \right.
\ee
which gives
\be
\mathcal G(x) = \frac{1}{|1-\bar x|^4}\ ,
\ee
namely that only the disconnected piece contributes to the 4-point function. This divergent part is then subtracted in perturbation theory and so the anomalous dimension of the $J$-currents vanishes, as expected.

All in all, we find
\be\label{gammas_ut_ii}
\gamma^{k\ell}= \frac{\lambda^2\pi^2}{2}\delta^{k\ell}\ .
\ee
which agrees with the result obtained in (\ref{gammas_ut}).

\subsection{Non-holomorphic states}\label{subsection_nholom}
Let us now compute the anomalous dimension of the non-holomorphic states of dimension (1,1), see eq. (\ref{NS_NS_states2}). We now need to go beyond the expressions in \cite{Gaberdiel:2015uca}. We define 
\be
\varphi_m := V_m\Big(\Psi^{I}_{-\frac12}\Psi^J_{-\frac12}\vac\Big)\ ,\qquad \tilde \varphi_m :=V_m\Big(\pr{1}_{-\frac12}\pr{2}_{-\frac12}\vac\Big)\ .
\ee
For the three states with the holomorphic part $\hat J^{3,\pm}$ (\ref{Jminus}), we find
\be
\mathcal G(x)= \sum_{m,n}||\varphi_{m} \tilde\varphi_{n}|O\rangle||^2 x^{-m-h} \bar x^{-n-\bar h}
= \left(\sum_{m}||\varphi_{m} |O\rangle||^2 x^{-m-h}\right) \left(\sum_{m}||\tilde\varphi_{m} |O\rangle||^2 \bar x^{-m-\bar h} \right)\ .
\ee
The holomorphic part works out just as before. For the anti-holomorphic part we have
\be
||\tilde\varphi_{-m} |\Phi^s\rangle||^2= m\ , \qquad m \geq 0
\ee
such that we get
\be
\sum_{m=0}^\infty m \bar x^{m-1}= \frac{1}{(1-\bar x)^2}\ .
\ee
In total we thus get the same correlation function as in the previous subsection, namely
\be
\mathcal G(x)= \left(  \frac{1}{(1-x)^2} + \frac{1}{2x}\right)
\frac{1}{(1-\bar x)^2}
\ee
which ensures that $\varphi$ gets lifted by the exact same amount $\textstyle{\frac{\lambda^2\pi^2}{2}}$.

For the 3 states in eq. (\ref{h1hb12ut}) with the holomorphic part $J^{3,\pm}$ one again finds that the anomalous dimension vanishes. All in all, we obtain
\be\label{gammas_ut_nh}
\gamma^{k\ell}= \frac{\lambda^2\pi^2}{2}\delta^{k\ell}\ .
\ee
This concludes the computations for the untwisted sector and shows that the holomorphic and non-holomorphic states are lifted up with the same amount.

\section{Lifting the twisted sector}\label{section_twist_lifting}

\subsection{Overview}\label{subsection_overview}
In the twisted sector, we have 16 fixed point sectors. A fixed point sector $\alpha$ is given by $[\alpha] \in \frac{1}{2}W/W$ where $W$ is the winding lattice. When convenient we can also label them by $\vec \epsilon \in \Z_2^4$. In each sector, there are 8 states of weight $(1,\textstyle\frac12)$, whose lifting matrix we need to investigate. In total there are $8\times 16 = 128$ states $|\varphi^i\rangle$. We will denote them by $|ij;\alpha\rangle$ and $|\bar i j;\alpha\rangle$, see eqs. (\ref{eighthone})-(\ref{NS_NS_twisted}). Here $i,j=1,2$ and $\alpha$ runs over the 16 fixed point sectors. These states are given by:
\be\label{h1states}
| i j;\alpha\rangle = \sqrt{2}\Xl{i}_{-1/2}\pl{j}_0|\sigma^{--}_\alpha\rangle\ , \qquad|\bar i j;\alpha\rangle = \sqrt{2}\Xbl{i}_{-1/2}\pl{j}_0|\sigma^{--}_\alpha\rangle\ .
\ee
Note that of those eight states per sector, only six are $\cN=4$ primary fields, see subsection \ref{subsection_twisted}. The remaining two are $G^{\alpha A}_{-1/2}$ descendants of the chiral primaries. In what follows, it is convenient not to distinguish between the two types of states. In the end of the day, we will of course find that only the primary fields get lifted, and that the descendants are protected from lifting.

Let us define the second order lifting matrix $D$. We want to compute the second order contribution to the two point functions of the NS-NS sector fields $\varphi^k$ of dimension $(1,\frac{1}{2})$
\be
\langle \varphi^{\ell\dagger}(z_1,\bar z_1)\varphi^k(z_4,\bar z_4) \rangle \ .
\ee
As before, the integral starts out as
\be\label{OOphiphi_herm}
\int d^2 z_2 d^2 z_3 
\langle \varphi^{\ell\dagger}(z_1,\bar z_1)\;\mathcal O(z_2,\bar z_2)\;\mathcal O(z_3,\bar z_3)\;\varphi^k(z_4,\bar z_4)\rangle\ ,
\ee
where $\mathcal O$ is the hermitian modulus defined in (\ref{ccmodulus_real}). The above integral contains 4 terms with different combinations of the complex moduli $O$ and $O^\dagger$ at positions $z_2$ and $z_3$. Let us for the moment consider the term
\be
\int d^2 z_2 d^2 z_3 
\langle \varphi^{\ell\dagger}(z_1,\bar z_1)O^\dagger(z_2,\bar z_2)O(z_3,\bar z_3)\varphi^k(z_4,\bar z_4)\rangle
\ee
We will analyze the other 3 terms later on in subsection \ref{subsection_WardContour}. Using eq. (\ref{2ndorder}), we find the regularized integral
\be
\pi\lambda^2\log\left(\frac{|z_1-z_4|^2}{\epsilon^2}\right) (z_1-z_4)^{-2}(\bar z_1-\bar z_4)^{-1} \int d^2 x  
\langle \varphi^{\ell\dagger}(\infty)O^\dagger(1)O(x)\varphi^k(0)\rangle\ .
\ee
We define the matrix
\be\label{DijGij}
D^{k\ell}\equiv-\int d^2 x \mathcal G^{k\ell}(x) :=-\int d^2 x   \langle \varphi^{\ell\dagger}(\infty,\infty)O^\dagger(1,1)O(x,\bar x)\varphi^k(0,0)\rangle
\ee
so that the actual lifting matrix is given by
\be\label{twistedliftingmatrix}
\gamma^{k\ell}= \pi \lambda^2 D^{k\ell}\ .
\ee

The matrix $D$ is a $128\times128$ matrix. We will see that it simplifies significantly. Let us first sketch the simplifications here and give the simplified form before discussing the details. Because the right-moving part of $\mathcal G^{k\ell}(x)$ has enough supersymmetry, as we will describe in section~\ref{subsection_WardContour}, we can use superconformal Ward identities to write it as a total derivative: $\mathcal G^{k\ell}(x) = -\partial_{\xb} I(x)$. Stokes' theorem (\ref{stokes}) allows to reduce the area integral to a contour integral 
\be\label{stokes_contours}
\int d^2 x\,\mathcal G^{k\ell}(x) \sim \oint_{0,1,\infty} dx 
I(x)
\ee
around the the insertion points $0,1,\infty$. Around 0 only the simple pole $x^{-1}\xb^0$ gives a contribution, and similarly for $1$ and $\infty$. Any other term either vanishes from the angular integral, or gives a power of $\epsilon$, which either goes to 0 or is regulated away if it diverges, as described in section~\ref{section_2ndorder}. This means that we only need to consider the OPE of $O(x)$ with the three other fields in $\mathcal G(x)$ in eq. (\ref{DijGij}), and in those OPEs we only need to keep track of fields with $(h,\bar h)=(1,0)$. This method is very similar to the one introduced in \cite{Gava:2002xb}, and further developed and used in \cite{deBoer:2008ss, Guo:2019pzk}.

By the usual argument $O$ has no marginal field in the OPE with itself, since otherwise its $\beta$ function does not vanish and so it would not be exactly marginal. It then follows that there is no contribution from $x=1$ to the contour integral (\ref{stokes_contours}). (We will give a detailed argument for this in section~\ref{ss:x1}.)

Concentrating on $x=0$, let us fix for concreteness $O$ to be in the fixed point sector $\alpha$. We then write
\be\label{D1D2}
D\equiv\Do \otimes \Dtw\ .
\ee
$\Do$ is a $16\times16$ matrix, encoding the information of the 16 fixed point sectors.  The only non-vanishing entry is the element $\Do_{\alpha\alpha}$:  for $x=0$, if $O$ is in the fixed point sector $\alpha$ and $\varphi_k$ in the fixed point sector $\beta$, then the fields in the OPE will be in the untwisted sector with momentum $\alpha+\beta$ \cite{Brunner:1999ce}. Note that unless this is in the vacuum sector, the ground state will have fractional conformal weight {\footnote{We are assuming that the torus does not have any enhanced symmetries.}}, so that it is impossible to obtain a state with $h=1$ . This is explained in more detail in appendix~\ref{app:bostwist}. The only states with $h=1$ can thus appear if $\alpha+\beta$ has vanishing momentum, that is if $\alpha+\beta=0$. The same holds true for $x=\infty$. Also note that due to momentum conservation $\mathcal G^{\beta\gamma}(x)$ vanishes unless $\beta=\gamma$. It thus follows that $\Do$ is diagonal, and that its only non-vanishing entry is the entry $\Do_{\alpha\alpha}$.

$\Dtw$ is the $8\times8$ matrix encoding the states (\ref{h1states}). We will see that $\Dtw$ simplifies even further: it is in fact block diagonal, with two $4\times4$ blocks $\Dth$, which are in turn again block diagonal
\be\label{D2}
\Dtw = \begin{pmatrix}\Dth&0\\0&\Dth\end{pmatrix}\ .
\ee
We will find in subsection \ref{lifting_matrix} that 
\be
\Dth = \frac{\pi}{2}
\begin{pmatrix}
	1&0&0&0\\0&1&0&0\\
	0&0&\frac12&-\frac12\\0&0&-\frac12&\frac12 
\end{pmatrix}\ ,
\ee
which has eigenvalues $\{\textstyle\frac{\pi}2,\textstyle\frac{\pi}2,\textstyle\frac{\pi}2,0\}$.

Diagonalizing $D$ thus leads to the following result: all the states outside of the fixed point sector  $\alpha$ are left invariant. In the fixed point sector $\alpha$, 2 out of the 8 states have eigenvalues 0 and are therefore left invariant. These states are exactly the $G^{-A}_{-1/2}$ descendants of the ground state and correspond to the modulus. The remaining 6 states have eigenvalue $\textstyle\frac{\pi}2$, and therefore (\ref{twistedliftingmatrix}) shows that they are lifted by the exact same amount, $\textstyle\frac{\lambda^2\pi^2}2$, as their partners in $U^{\tilde l=1/2}_1$ (\ref{gammas_ut_ii}) and $T^{\tilde l=0}_1$ (\ref{gammas_ut_nh}), combining into a long representation of the small $\cN=4$ algebra. In terms of (\ref{h1states}), the 2 moduli correspond to the states
\be\label{2nonlifting}
\textstyle\frac1{\sqrt2}(|\bar 1 1\rangle + |\bar 2 2\rangle) \ , \qquad\textstyle\frac1{\sqrt2}(|21\rangle - |12\rangle)\ ,
\ee
and the 6 lifted states are
\be\label{6lifting}
|\bar 1 2\rangle\ , \ |\bar 2 1\rangle\ , \ |11\rangle\ , \ |22\rangle\ ,\ \textstyle\frac{1}{\sqrt{2}}(|\bar 11\rangle - |\bar 22\rangle)\  ,\
\textstyle\frac{1}{\sqrt{2}}(|21\rangle + |12\rangle)
\ee
where we have suppressed the index $\alpha$, as we will for the rest of this section (see also subsection \ref{subsection_twisted}).

\subsection{Contour integrals}\label{subsection_WardContour}
We work out the above claims in detail in the remainder of this section. In principle, we now need to evaluate the integral (\ref{DijGij}) over $dx d\xb$. We note, however, that the right-movers are all BPS states. This allows us to apply right-moving Ward identities, which reduce the partition function to a total derivative.

We will take $|\varphi\rangle$ to be states (\ref{h1states}). Note that they are of the form $|\varphi\rangle=\mathcal{A}|\sigma^{--}\rangle$, where $\mathcal{A}$ is a bilinear in purely left-moving descendants. To obtain their lifting, we want to compute the 4-point function:
\be\label{OOphiphi_i}
\mathcal I_{\varphi}\equiv
\langle\varphi^k(z_1,\bar z_1)\;\;O(z_2,\bar z_2)\;\;O^{\dagger}(z_3,\bar z_3)\;\;\varphi^{\ell\dagger}(z_4,\bar z_4)\rangle\ .
\ee
We use the Ward identities discussed in appendix \ref{app_ward} and simplify the 4-point function.

Using eq. (\ref{ccmodulus}) for the modulus and the Ward identity (\ref{ward_G_contour}), we write
\be\label{ward_O}
O(z_2,\bar z_2)=
\oint_{\bar z=\bar z_2}\frac{d\bar z}{2\pi i}\;\frac{\bar z-\bar z_1}{\bar z_2-\bar z_1}\;
\tilde G^{+1}(\bar z)\Big(G^{+1}_{-\frac12}\sigma^{--}\Big)(z_2,\bar z_2)\ .
\ee
Inserting this in eq. (\ref{OOphiphi_i}) we have
\be\label{OOphiphi_ii}
\mathcal I_{\varphi}=
\Big\langle\varphi^k(z_1,\bar z_1)\oint_{\bar z=\bar z_2}\frac{d\bar z}{2\pi i}\;\frac{\bar z-\bar z_1}{\bar z_2-\bar z_1}\;
\tilde G^{+1}(\bar z)\Big(G^{+1}_{-\frac12}\sigma^{--}\Big)(z_2,\bar z_2)\;O^\dagger(z_3,\zb_3)\;\varphi^{\ell\dagger}(z_4,\bar z_4)\Big\rangle\ .
\ee
Next, we deform the contour around the other insertion points. Noting that $|\varphi\rangle=\mathcal{A}|\sigma^{--}\rangle$ and using the OPEs between the $G$'s and the (anti-)chiral primaries
\be\label{GphiOPE}
\tilde G^{+1}(\bar z)\sigma^{--}(w,\bar w)\sim\frac{ (\tilde G^{+1}_{-\frac12}\sigma^{--})(w,\bar w)}{\bar z-\bar w}\ ,\qquad
\tilde G^{+1}(\bar z)\sigma^{++}(w,\bar w)\sim0\ ,
\ee
we find that, due to the choice of the function in (\ref{ward_O}), neither $\zb_1$ nor $\zb_4$ contribute.

The only contribution then comes from the modulus $ O^\dagger(z_3,\zb_3)$. The OPE between the various moduli are given by
\bea\label{GO_OPE}
&&\!\!\!\!\!\!\!\!\!\!\!\!\!\!\!\!\!\!\!\!
\frac12G^{+A}(z)\;\Big(G^{-B}_{-\frac12}\cdot\sigma^{+}(w)\Big)\sim\epsilon^{AB}\partial_w\Big(\frac{\sigma^{+}}{z-w}\Big)\ ,\quad
\frac12G^{-A}(z)\;\Big(G^{-B}_{-\frac12}\cdot\sigma^{+}(w)\Big)\sim0\ ,\\
&&\!\!\!\!\!\!\!\!\!\!\!\!\!\!\!\!\!\!\!\!
\frac12G^{-A}(z)\;\Big(G^{+B}_{-\frac12}\cdot\sigma^{-}(w)\Big)\sim\epsilon^{AB}\partial_w\Big(\frac{\sigma^{-}}{z-w}\Big)\ ,\quad
\frac12G^{+A}(z)\;\Big(G^{+B}_{-\frac12}\cdot\sigma^{-}(w)\Big)\sim0\ ,\nonumber
\eea
and likewise for the right-moving fields. We therefore find that
\bea\label{OOphiphi_iii}
\mathcal I_{\varphi}=-\partial_{\bar z_3}\Bigg(\frac{\bar z_3-\bar z_1}{\bar z_2-\bar z_1}\;
\Big\langle \varphi^k(z_1,\bar z_1)\;O'(z_2,\bar z_2)\;O^{'\dagger}(z_3,\bar z_3)\;\varphi^{\ell\dagger}(z_4,\bar z_4)\Big\rangle\Bigg)\ .
\eea
Here we have introduced the notation
\be\label{half}
O'\equiv G^{+1}_{-\frac12}\sigma^{--}
\ee
that is the operator with the same left-moving structure as that of the modulus. Note that now the right moving sector of this correlation function \emph{only} contains ground states and no descendants. Similarly, by using the analogue of (\ref{ward_O}) for $O^\dagger$ we can also write (\ref{OOphiphi_i}) as
\bea\label{OOphiphi_iii2}
\mathcal I_{\varphi}=-\partial_{\bar z_2}\Bigg(\frac{\bar z_2-\bar z_4}{\bar z_3-\bar z_4}\;
\Big\langle \varphi^k(z_1,\bar z_1)\;O'(z_2,\bar z_2)\;O^{'\dagger}(z_3,\bar z_3)\;\varphi^{\ell\dagger}(z_4,\bar z_4)\Big\rangle\Bigg)\ .
\eea

Let us now use (\ref{OOphiphi_iii}) and (\ref{OOphiphi_iii2}) to express our 4-point functions $\mathcal G(x)$ in eq. (\ref{DijGij}). For the term
\be
\mathcal G^{k\ell}(x)= \langle\varphi^\ell| O'^{\dagger}(1)O'(x)|\varphi^k\rangle
\ee
we use (\ref{OOphiphi_iii2}) and send the $z_i$ to their respective positions $z_1\to0$, $z_2\to x$, $z_3\to 1$, and $z_4\to \infty$ to get
\be\label{I1}
\mathcal G^{k\ell}(x)= -\partial_{\xb}\langle \varphi^\ell|O'^\dagger(1)O'(x)|\varphi^k\rangle=:-\partial_{\xb} I_1(x)\ .
\ee

We now consider the other 3 combination of the moduli in eq. (\ref{OOphiphi_herm}). For the term
\be
{\mathcal G}^{k\ell}(x)\equiv\langle\varphi^\ell| O'(1)O'^\dagger(x)|\varphi^k\rangle
\ee 
we use (\ref{OOphiphi_iii}) instead, with $z_2\to 1,z_3\to x$, which then gives
\be\label{I2}
{\mathcal G}^{k\ell}(x)= -\partial_\xb\left( \xb \langle \varphi^\ell|O'(1)O'^\dagger(x)|\varphi^k\rangle \right) =: -\partial_\xb I_2(x)\ .
\ee
Finally, for the cases with 2 $O'$'s or 2 $O'^\dagger$'s in the deformation integral, the 4-point function vanishes due to conservation of the R-charge and Wick contraction of bosonic fields. Note also that the other hermitian combination of the moduli, $\widehat{\mathcal O}$, in eq. (\ref{ccmodulus_real}) yields the same result for the mixing matrix.

We use Stokes' theorem in both cases (\ref{I1}) and (\ref{I2}) to reduce the integral to a contour integral around $0,1,\infty$. 
\be
\frac{i}{2}\int dx d\xb\,\mathcal G^{k\ell}(x) =-\frac{i}{2}\oint_{0,1,\infty} dx I_{1,2}(x) \ .
\ee
We evaluate the contour integral by evaluating the OPE of $O'(x)$ with the three other fields at $0,1,\infty$. At $x=0$, for instance, we have
\be\label{Opvarphi}
O'(x)\varphi(0) \sim\frac{1}{x^{2-h_\phi}}\,\frac{1}{\xb^{1-\bar h_\phi}}\,\phi(0) \ .
\ee
Parametrizing $x=\epsilon e^{i\theta}$, we see that the only constant contribution comes from the term $x^{-1}$. For (\ref{I1}) this means that $\phi$ is marginal, whereas for (\ref{I2}) it means that $\phi$ has weight $(1,0)$. In any other case we either get 0, or a power of $\epsilon$, which either goes to 0 or is regulated away if it diverges, as described in section~\ref{section_2ndorder}.

\subsection{Contribution from $x=1$}\label{ss:x1}

Let us now discuss the contributions from the three different contours. We first argue that there is no contribution from $x=1$. This is essentially the argument that exactly marginal fields cannot have marginal fields in their OPE, since otherwise they would no longer be marginal at first order. We are essentially repeating the argument of \cite{Dixon:1987bg, Gukov:2004ym} here, 
but concentrate on the left-movers only.

We use the Ward identities to compute
\be
\langle O'(z_1)O^{'\dagger}(z_2)\phi(z_3) \rangle\ .
\ee
Assume that $\phi$ is a field with $h=1$. It has to be uncharged under the $SU(2)$ R-symmetry. The OPE of $G^{-A}(z)$ with $\phi$ is
\be
G^{-A}(z)\phi(0) \sim \frac{1}{z^2}V\big(G^{-A}_{\frac12}|\phi\rangle,0\big) + \frac{1}{z}V\big(G^{-A}_{-\frac12}|\phi\rangle,0\big)+\ldots\ .
\ee
Note that the state $G^{-A}_{1/2}|\phi\rangle$ has $h=\textstyle\frac12$ and $q=\pm\textstyle\frac12$, and is therefore a (anti-)chiral primary. This means that the only way to have a double pole in the OPE is if $\phi$ is the $G^{-A}_{-1/2}$ descendant of a chiral primary. This is, however, the argument in \cite{Dixon:1987bg, Gukov:2004ym} which shows that the 3-point function vanishes in this case. Without loss of generality we will therefore assume that the OPE only contains a simple pole.

We then use the Ward identities to compute the 3-point function as
\bea
&&\langle O'(z_1)O^{'\dagger}(z_2)\phi(z_3) \rangle =\oint_{z_1}dz\,\frac{z-z_3}{z_1-z_3}\,\Big\langle G^+(z)\sigma^{--}(z_1) O^{'\dagger}(z_2)\phi(z_3)\Big\rangle\\
&&\qquad\qquad\qquad\qquad\;\;\;\,
= - \oint_{z_2}dz \partial_{z_2}\bigg(\frac{z-z_3}{z_1-z_3}\,\frac{1}{z-z_2}\,\langle \sigma^{--}(z_1) \sigma^{++}(z_2)\phi(z_3)\rangle\bigg)\nonumber\\
&&\qquad\qquad\qquad\qquad\quad
= \partial_{z_2}\bigg(\frac{z_2-z_3}{z_1-z_3}\langle\sigma^{--}(z_1) \sigma^{++}(z_2)\phi(z_3)\rangle\bigg)
=\partial_{z_2}\frac{C}{(z_1-z_3)^2}= 0\ .,\nonumber
\eea
where $C$ is a constant. In the second line we deformed the contour away from $z_1$ and used that the $G^{-A}(z)\,\phi(z_3)$ OPE only has a simple pole, see eq. (\ref{GphiOPE}). This pole is canceled by the factor in the numerator, so that the contour around $z_3$ does not give a contribution. The integrand decays fast enough at infinity so that the only contribution comes from $z_2$, where we used the usual OPE of two moduli (\ref{GO_OPE}). This establishes that the 3-point function vanishes, and that there is indeed no contribution from $x=1$.

\subsection{Contributions for $x=0$ and $x=\infty$}

Let us now evaluate the contour integral of $I_{1,2}(x)$ around $x=0$, see eqs. (\ref{I1}) and (\ref{I2}). We expand them as
\be
I_{1,2}(x)= \sum_{a,b}I^{a,b}_{1,2}\,x^a \xb^b\ .
\ee
By the argument below eq. (\ref{Opvarphi}), the only term that contributes is $x^{-1}\xb^0$. Let us first discuss the right-moving part. For the right-moving fermionic part of $I_1(x)$ (\ref{I1}), we use (\ref{feranti2}) to get the contribution
\be
\langle\tilde\sigma^-_f| 
(\tilde\sigma^-_f)^\dagger(1)\,\tilde\sigma^-_f(x)|\tilde\sigma^-_f\rangle= \frac{\xb^{\frac12}}{(1-\xb)^{\frac12}}\ .
\ee
For $I_2(x)$ (\ref{I2}), we use (\ref{feranti1}) to get
\be
\xb\cdot \langle\tilde\sigma^-_f| \tilde\sigma^-_f(1)\,(\tilde\sigma^-_f)^\dagger(x)|\tilde\sigma^-_f\rangle
 = \frac{\xb^{\frac12}}{(1-\xb)^{\frac12}}\ ,
\ee
where for convenience we have included the additional factor $\xb$ in the definition of the anti-holomorphic fermionic contribution. Next consider the bosonic part: a bosonic state in the intermediate channel with anti-holomorphic weight $h_R$ gives contribution $\xb^{-1/2+h_R}$, so that in total we get
\be
\sim  \frac{\xb^{h_R}}{(1-\xb)^{\frac12}} = \xb^{h_R} + \ldots\ .
\ee
It follows that the only state giving a constant term contribution is for $h_R=0$, that is for the vacuum. The overall right-moving contribution is thus simply 1.

Next, consider the left-moving part. For the $\xb^0$ term, we simply have
\be
I(x)= \sum_{\phi} x^{-2+h_\phi} C_{\varphi^{\ell\dagger} O \phi^\dagger }  C_{\phi O^\dagger\varphi^k }\ ,
\ee
where $C_{\phi O\varphi}$ are 3-point functions and the sum is over all the primary fields and their descendants running in the internal channel. The contour integral vanishes for all terms except for $\textstyle\frac1x$. Combining all contributions, we immediately find that $h_\phi=1$. We can therefore concentrate on such states. Our claim is that the only such states that can give non-vanishing contributions are 
\be\label{h1internal}
|\phi^{IJ}\rangle\equiv\psi^{(I)}_{-\frac12}\psi^{(J)}_{-\frac12}|0\rangle\ .
\ee
Note that these can also be written as $J_{-1}$ descendants of the vacuum for the R-symmetry and flavor symmetry currents $\hat J$. To see that these are the only states that can contribute, note that $\Xl{i}_{-1}\vac$ has vanishing 3-point functions, since $\varphi$ and $\Phi$ have a single bosonic descendant, so that we have three bosonic modes.

Similarly, we exclude a single fermionic descendant of a $h=\textstyle\frac12$ winding-momentum state. The only remaining possibility is thus that we have a winding-momentum mode ground state $|1,0\rangle$. We will exclude that possibility by assuming that the torus is not at the self-dual radius.  Note that the vacuum can only appear if the two fields are in the same fixed point sector. This confirms our statements in subsection \ref{subsection_overview}, see below eq. (\ref{D1D2}).

All we need to do now is thus to compute all 3-point functions with fields (\ref{h1internal}):
\be
C_{\phi O^\dagger\varphi^k } = \langle \phi| O^\dagger(1) |\varphi^k\rangle\ .
\ee
The total contribution from $x=0$ is then simply
\be\label{totcont0}
\frac{i}{2}\oint_{0}dx I(x) =  \pi \sum_{h_\phi=1}C_{\varphi^{\ell\dagger} O\phi^\dagger }C_{\phi O^\dagger\varphi^k }=
\pi \sum_{h_\phi=1}  C_{\phi O^\dagger\varphi^\ell }^*C_{\phi O^\dagger\varphi^k }
\ee
The same argument goes through for $x=\infty$. The only difference is that we need to exchange $O$ and $O^\dagger$, since $O^\dagger$ now approaches $\infty$. We then find that eq. (\ref{DijGij}) reads
\be\label{totcont}
D^{k\ell}= \pi \sum_{h_\phi=1}  C_{\phi O^\dagger\varphi^\ell}^*C_{\phi O^\dagger\varphi^k}+
\pi \sum_{h_\phi=1}  C_{\phi O\varphi^\ell}^*C_{\phi O\varphi^k}
\ee
Note that this also implies that we do not need to work with the actual hermitian modulus $\frac{1}{\sqrt{2}}(O+O^\dagger)$, as the result is exactly the same.

\subsection{3-point functions and the lifting matrix}\label{lifting_matrix}
Let us now compute the 3-point functions. As mentioned above, the 8 states naturally decompose into two sets $|ij\rangle$ and $|\bar ij\rangle$, see eq. (\ref{h1states}). We will see that the mixing matrix $D$ (\ref{totcont}) is block diagonal with respect to these sets. Note that all the 3-point function have two twist fields and one untwisted field. Their computation is fairly straightforward: the bosonic part consists of only two bosonic descendants of the twist fields and the untwisted vacuum. It is thus simply a 2-point function of form
\be\label{boson2pf}
\langle \sigma|\Xl{i}_{1/2} \Xl{j}_{-1/2}|\sigma\rangle=\frac{\delta^{ij}}{2}\ .
\ee
The fermionic part can be computed using the bosonization techniques described in section \ref{spinfields}, see also appendix \ref{app:fermtwist}.

First consider the 3-point functions for $|ij\rangle$, 
\be
C_{\phi\mathcal O^{\prime\dagger}\varphi}=\langle\phi| O^{\prime\dagger}(1)|ij\rangle
\ee
We will obtain them from the full 3-point function
\be\label{phi_odagger_varphi1}
\Big\langle\phi^\dagger(y)\;\;(\partial{X}^1_{-\frac12}\bar\Psi^1_0+\partial{X}^2_{-\frac12}\bar\Psi^2_0)\sigma^{+\dot+}(z)\;\;
\sqrt2\partial X^i_{-\frac12}\Psi^j_{0}\sigma^{-\dot-}(w)\Big\rangle
\ee
where $\phi(y)$ is a purely fermionic operator corresponding to a state (\ref{h1internal}). Note that we are only keeping track of the left-moving part, since we already established that the anti-holomorphic part contributes 1. We next evaluate the bosonic contribution. Since $\phi(y)$ is a purely fermionic operator, the bosonic part is simply a 2-point function of the form (\ref{boson2pf}). In the case at hand it vanishes, as there are no bosonic contractions possible. It then follows that
\be\label{nonbarstates}
\langle\phi|\mathcal O^{\prime\dagger}(1)|ij\rangle=0\ ,\quad\rm{for}\quad|11\rangle\ ,\quad \ |22\rangle\ ,\quad |21\rangle\ ,\quad  |12\rangle \ .
\ee

We next consider the states  $|\bar ij\rangle$:
\be\label{barstates}
|\bar11\rangle\ ,\quad \ |\bar22\rangle\ ,\quad |\bar21\rangle\ ,\quad  |\bar12\rangle\ .
\ee
Here the 3pt function is of the form
\be\label{phi_odagger_varphi2}
\Big\langle\phi^\dagger(y)\;\;(\partial{X}^1_{-\frac12}\bar\Psi^1_0+\partial{X}^2_{-\frac12}\bar\Psi^2_0)\sigma^{+\dot+}(z)\;\;
\sqrt2\partial\bar X^i_{-\frac12}\Psi^j_{0}\sigma^{-\dot-}(w)\Big\rangle\ .
\ee
Note that now there are allowed Wick contraction for the bosons, so that  the 3-point function may be non-vanishing. Let us start with the external state $|\bar 2 1\rangle$ and evaluate the correlation function for $\langle \phi^{\bar 2 1}| O^\dagger(1) |\bar 2 1\rangle$. We have $\phi^{\bar 2 1}(y)=\;\no{e^{i(H^1-H^2)}}(y)$, and we need to consider
\bea\label{OOvarphivarphi_iii}
&&\sqrt2\,\Big\langle \no{e^{i(-H^1+H^2)}}(y)\;\;\no{(e^{\frac i2(-H^1+H^2)}{ X}^1_{-\frac12}+e^{\frac i2(H^1-H^2)}\partial{ X}^2_{-\frac12})\sigma_b}(z)\times\nonumber\\
&&\qquad\qquad\qquad\qquad\qquad\qquad\qquad\qquad\qquad\;\;\;
\times\no{e^{\frac i2(H^1-H^2)}\partial\bar{ X}_{-\frac12}^2 \sigma_b}(w)\Big\rangle\nonumber\\
&&= \frac{1}{\sqrt{2}}\frac{1}{(z-w)^{3/2}}\Big\langle :e^{i(-H^1+H^2)}:(y):e^{\frac i2(H^1-H^2)}:(z):e^{\frac i2(H^1-H^2)}:(w)\Big\rangle\nonumber\\
&&= \frac{1}{\sqrt{2}}\,\frac{1}{(z-w)^{3/2}}\,\frac{(z-w)^{\frac12}}{(y-z)(y-w)}\ .
\eea
Here in the third line we evaluated the bosonic contribution to the correlation function, which is simply the two point function $\langle\sigma_b|\Xl{2}_{1/2}\Xbl{2}_{-1/2}|\sigma_b\rangle$, and in the last line we evaluated the fermionic 3-point function. From this we can immediately read off
\be\label{Odaggerbar21}
\langle \phi^{\bar 2 1}| O^\dagger(1,1) |\bar 2 1\rangle =\frac{1}{\sqrt{2}}\ .
\ee

We also find that $\langle \phi| O^\dagger(1) |\bar 2 1\rangle=0$ for any of the other five $\phi$'s in (\ref{h1states}), simply because the fermionic part does not satisfy R-charge conservation unless $\phi=\phi^{\bar 21}$.

By the exact same computation we find for $|\bar 1 2\rangle$:
\be\label{Odaggerbar12}
\langle \phi^{\bar 1 2}| O^\dagger(1,1) |\bar 1 2\rangle =\frac{1}{\sqrt{2}}\ ,
\ee
and vanishing 3-point function for any other $\phi$'s.

Next consider $\langle \phi^{\bar 1 1}| O^\dagger(1)|\bar 11\rangle$. Note that now $\phi^{\bar 1 1}(y)=i\partial H^1(y)$. The relevant 3-point function is thus
\begin{multline}\label{C3pt1b1}
\frac{1}{\sqrt{2}}\frac{1}{(z-w)^{3/2}}
\Big\langle i\partial H^1(y)
:e^{\frac i2(-H^1+H^2)}:(z):e^{\frac i2(H^1-H^2)}:(w)\Big\rangle\\
= \frac{1}{\sqrt{2}}\,\frac{1}{(z-w)^{3/2}}\left(\frac{-\frac12}{y-z}+\frac{\frac12}{y-w}\right)
\frac{1}{(z-w)^{1/2}}\qquad\qquad\qquad\qquad\\
= - \frac{1}{2\sqrt{2}}\,\frac{1}{(y-z)(y-w)(z-w)}\qquad\qquad\qquad\qquad\qquad\qquad\qquad\qquad\;
\end{multline}
where we have used (\ref{OPExexpx}) to do pole subtraction. This yields
\be
\langle \phi^{\bar 1 1}| O^\dagger(1)|\bar 11\rangle= -\frac{1}{2\sqrt{2}}\ .
\ee
The only other non-vanishing 3-point function for $|\bar 11\rangle$ is $\langle \phi^{\bar 22}| O^\dagger(1)|\bar 11\rangle$. Now $\phi^{\bar 22}(y)=i\partial H^2(y)$, so that the same computation as in (\ref{C3pt1b1}) gives a minus sign,
\be
\langle \phi^{\bar 22}| O^\dagger(1)|\bar 11\rangle= \frac{1}{2\sqrt{2}}\ .
\ee
All other $\phi$'s give vanishing 3-point functions due to charge conservation.

Finally consider $\langle \phi^{\bar 1 1}| O^\dagger(1)|\bar 22\rangle$. The computation is the same as above, except that now we get
\begin{multline}\label{C3pt2b2}
\frac{1}{\sqrt{2}}\frac{1}{(z-w)^{3/2}}
\Big\langle i\partial H^1(y)
:e^{\frac i2(H^1-H^2)}:(z)
:e^{\frac i2(-H^1+H^2)}:(w)\Big\rangle\\
=  \frac{1}{2\sqrt{2}}\frac{1}{(y-z)(y-w)(z-w)}\ ,\qquad\qquad\qquad\qquad
\end{multline}
so that 
\be
\langle \phi^{\bar 1 1}| O^\dagger(1)|\bar 22\rangle
=  \frac{1}{2\sqrt{2}}\ .
\ee
Similarly we obtain 
\be
\langle \phi^{\bar22}| O^\dagger(1)|\bar 22\rangle
=  -\frac{1}{2\sqrt{2}}\ .
\ee

We can then immediately compute the contribution at $x=\infty$, that is the term $C_{\phi O\varphi^j }$ in (\ref{totcont}). For this we can simply use
\be
C_{\phi O\varphi}= C^*_{\varphi^\dagger O^\dagger \phi^\dagger}
=- C^*_{\phi^\dagger O^\dagger\varphi^\dagger }
\ee
and note that under taking the adjoint the states $|\bar i j\rangle$ get mapped to states of the form $|ij\rangle$, for which we have already shown that $C_{\phi O^\dagger\varphi}$ vanishes. It follows that $C_{\phi O\varphi^j }=0$, so that there is no contribution from $x=\infty$. It then also immediately follows from (\ref{totcont}) that there are no matrix elements between (\ref{nonbarstates}) and (\ref{barstates}). This establishes the block diagonal form (\ref{D2}) of $\Dtw$. Collecting all 3-point functions and using (\ref{totcont}), we find that the the upper block of the mixing matrix (\ref{D2}), associated with the states $|\bar ij\rangle$, only contains contributions from the $x=0$ limit and is of the form
\be\label{mixing_block}
\Dth = \frac{\pi}{2}
\begin{pmatrix}
	1&0&0&0\\0&1&0&0\\
	0&0&\frac12&-\frac12\\0&0&-\frac12&\frac12 
\end{pmatrix}\ .
\ee
We can now make a change of basis to more clearly exhibit the difference between moduli and primary fields by using eigenvectors of $\Dth$. Namely, $\frac{1}{\sqrt{2}}(|\bar 11\rangle -|\bar 2 2\rangle)$ is an eigenvector with eigenvalue $\frac\pi2$. On the other hand, $\frac{1}{\sqrt{2}}(|\bar 11\rangle +|\bar 2 2\rangle)$ is an eigenvector with eigenvalue 0, see eqs. (\ref{2nonlifting}) and (\ref{6lifting}). This confirms indeed that descendant corresponding to the modulus is not lifted, as expected.

We shall repeat this computation for the other 4 states of the form $|ij\rangle$: in this case the 3-point function $\langle\phi|O'(1)|ij\rangle$ at $x=0$ is non-vanishing whereas $\langle\bar i\bar j|\mathcal O^{\prime}(1,1)|\phi\rangle$ vanishes. The lower block of the mixing matrix (\ref{D2}) is again found to be of the form (\ref{mixing_block}). Now $\frac{1}{\sqrt{2}}(|21\rangle - |12\rangle)$ is the state that gets lifted, and $\frac{1}{\sqrt{2}}(|21\rangle + |12\rangle)$ corresponds to the unlifted state. All in all, we find that the mixing matrix $D^{(2)}$ is of the form (\ref{D2}).

\section{Symmetry surfing}\label{s:symmetrysurfing}
\subsection{Perturbing in a general direction}
So far we have fixed a specific fixed point sector $\alpha$, and chosen $O$ to lie in that sector. Let us now discuss what happens if if perturb in a general direction, that is if we choose $O$ to be a linear combination of the $O_\alpha$. For convenience, let us use the actual hermitian modulus $\mathcal O_\alpha$ defined in eq. (\ref{ccmodulus_real}). We take a general modulus as a (real) linear combination of the fixed point sectors
\be\label{generaldirection}
\cO = b^{\beta}\cO_\beta\ ,
\ee
where $b^\alpha$ is a unit vector, $b^\alpha b_\alpha=1$.

When working with (\ref{generaldirection}), clearly our computations for the untwisted states does not change. For the twisted states, the matrix $\Dtw$ does not change from our previous computation. The matrix $\Do$ however will be a linear combination of the previous computations. To this end, consider 
\be
\langle \varphi_\alpha|\cO_\beta(1)\cO_\gamma(x)|\varphi_\delta \rangle\ .
\ee
Let us proceed to evaluate this correlator in the same way as we did above. By the argument in section~\ref{ss:x1}, there is no contribution from $x=1$. The OPE of $\cO_\gamma$ at $x=0$ can only give a non-vanishing contribution if $\gamma=\delta$, since otherwise the vacuum will not appear. Similarly, we can only get a contribution at $x=\infty$ if $\gamma=\alpha$. Charge conservation then immediately implies that $\beta=\alpha$ or $\beta=\delta$. Putting this together means that
\be
\Do_{\alpha\delta} = \frac12(\delta_{\alpha\beta}\delta_{\gamma\delta}
+\delta_{\alpha\gamma}\delta_{\beta\delta})\ .
\ee
This leads to the lifting matrix
\be
\Do_{\alpha\delta}= b_\alpha b_\delta\ .
\ee
From this we immediately see that $\Do$ has an eigenvector $b_\alpha$ with eigenvalue 1. This means that the states parallel to the direction of perturbation get lifted. The 15 eigenvectors orthogonal to $b_\alpha$ have eigenvalue 0, so that these states do not get lifted.

\subsection{Symmetry surfing}
Let us now discuss the implications of this for Mathieu moonshine. Our results show that by perturbing in any direction, we can pick out a 90 dimensional subspace of $\textstyle\frac14$-BPS states with $h=1$. It is of course tempting to suggest that these states will form part of the Mathieu moonshine module. The more challenging part, however, is to find the action of $\mg$. The symmetry surfing proposal of \cite{Taormina:2013jza, Taormina:2013mda} makes a proposal for how to construct the action of maximal subgroup $G$ of $\mg$, the octad group. This subgroup is constructed by `surfing' on the sublocus of Kummer surfaces to reach three special tori with enhanced symmetry, whose geometric symmetries are combined into $G$. These geometric symmetries permute the fixed point sectors. If we want  $G$ to act on 90, we need to make sure that it is a $G$-invariant subspace. This means that we need to pick the modulus to be invariant under permutations of the fixed point sectors, which is leads us to consider
\be\label{TW}
\cO^s:=\frac{1}{4}\sum_{[\alpha] \in \frac{1}{2}W/W}\cO_\alpha\ .
\ee

To obtain the modulus (\ref{TW}) we take $b=\frac14(1,1,\ldots,1)$, giving
\be
\Do = \frac1{16}\begin{pmatrix} 1 & 1 &\cdots &1\\
	1 & 1 &\cdots& 1\\
\vdots & \vdots &\ddots &\vdots\\
1 & 1 & \cdots &1 \end{pmatrix}\ ,
\ee 
where the factor $\textstyle\frac1{16}$ comes from the proper normalization of $\cO^s$. This matrix has one eigenvalue $1$, corresponding to the eigenvector $O^s$, and 15 eigenvalues 0. Again, this is exactly what we expect for the symmetry surfing story: namely, the states corresponding to the linear combination $\cO^s$ get lifted, whereas the 15 directions orthogonal to it do not, just as suggested in \cite{Gaberdiel:2016iyz}. It would be interesting to repeat our analysis for the higher $\textstyle\frac14$-BPS states and check if it agrees with the pattern proposed in \cite{Gaberdiel:2016iyz}.

\section*{Acknowledgments}
We are happy to thank Shouvik Datta, Eric D'Hoker, Lorenz Eberhardt, Daniel Friedan, Matthias Gaberdiel, Lothar G\"ottsche, Martin Raum, Anne Taormina, and Katrin Wendland for helpful discussions and correspondence. We thank Matthias Gaberdiel for comments on the manuscript. CAK thanks Greg Moore, Hirosi Ooguri, and Natalie Paquette for discussions and collaboration on a related project. We thank the referee for helpful comments. CAK and IGZ were supported by the Swiss National Science Foundation through the NCCR SwissMAP for part of this research.

\appendix

\section{The $\cN=(4,4)$ SCA for $\mathbf{T}^4/\Z_2$}\label{appendix:spectrum}
\label{app:N4}
We denote the complex fermions and complex bosons of $\mathbb{T}^4$
as 
\be
\pl{i} \ , \quad \pbl{i}  \ , \qquad  \Xl{i}\ , \quad \Xbl{i}  \ ,
\ee
where $i=1,2$. We will also sometimes use uppercase indices to write
\be
\pl{I}\ ,\qquad I=1,2,3,4
\ee
with the understanding that $\pl{3}:=\pbl{1}$ and $\pl{4}:=\pbl{2}$. In terms of these complex fields, the R-symmetry $\mathfrak{su}(2)$ currents are
\begin{eqnarray}\label{J}
J^{3} & = & \frac{1}{2} \Bigl(  \pl{1} \pbl{1} + \pl{2} \pbl{2}\Bigr) = \frac{1}{2i} \delta_{ij}\pl{i}\pbl{j}\ ,\\[2pt]
J^{+} & = & - \pl{1} \pl{2} = -\frac{1}{2}\epsilon_{ij}\pl{i}\pl{j}\nonumber\ ,\\[2pt]
J^{-} & = & \pbl{1} \pbl{2} =\frac{1}{2}\epsilon_{ij}\pbl{i}\pbl{j}\ ,\nonumber
\end{eqnarray}
where we work in the usual Cartan-Weyl basis for $\mathfrak{su}(2)$, and the bilinears are appropriately normal ordered.

Next we define the four supercurrents $G^{\alpha A}$:
\begin{eqnarray}\label{Gs}
G^{+1} = \delta_{ij}\pl{i}\Xbl{j} && G^{-1} = \epsilon_{ij}\pbl{i}\Xbl{j}\\
G^{+2} = -\epsilon_{ij}\pl{i} \Xl{j} && G^{-2} = \delta_{ij}\pbl{i}\Xl{j}\nonumber
\end{eqnarray}
The index $\alpha=\pm$ indicates the R-charge, that is the $G^{aA}$ form two doublets under the ${{su}}(2)$  R-symmetry.

Note that there is also a second  `flavor' ${SU}(2)$ symmetry which is the outer automorphism of the small $\cN=4$ superconformal algebra. The corresponding currents are given by
\begin{eqnarray}\label{Jminus}
\hat J^{3} & = & \frac{1}{2i} \Bigl( - \pl{1} \pbl{1} + \pl{2} \pbl{2} \Bigr)\ ,\\[2pt]
\hat J^{+} & = & - \pl{1} \pbl{2} \label{J-}\ ,\nonumber \\[2pt]
\hat J^{-} & = & \pbl{1} \pl{2} \ .\nonumber
\end{eqnarray}

\section{Twisted correlation functions}\label{app_4pfs}
\subsection{Bosonic twist fields}\label{app:bostwist}
Let us summarize the result for bosonic twist fields as described in \cite{Dixon:1986qv}. For a single boson on $S^1/\Z_2$, the 4-point function of four twist fields $\sigma_\alpha$ is given by
\be\label{G4twists}
G_{\e_0,\e_1}(x):=
\langle \sigma_{0}(\infty)\sigma_{ \epsilon_1}(1)\sigma_{ \epsilon_1+ \epsilon_0}(x)\sigma_{\epsilon_0}(0) \rangle
=2^{-\frac23}\frac{1}{|x(1-x)|^{\frac1{12}}} Z_{\e_0,\e_1}(\tau)
\ee
where
\be\label{Zoneboson}
Z_{\e_0,\e_1}(\tau)=\frac{1}{|\eta(\tau)|^2}\sum_{m\in \Z,n\in 2\Z+\epsilon_0}(-1)^{m\epsilon_1} 
q^{\frac14(\frac mR+\frac{nR}2)^2}\bar q^{\frac14(\frac mR-\frac{nR}2)^2}
\ee
and $q=e^{2\pi i\tau}$ is related to $x$ by
\be
x(\tau) = \frac{\theta_4^4}{\theta_3^4}= 
\prod_{n=1}^\infty \left(\frac{1-q^{n-\frac12}}{1+q^{n-\frac12}}\right)^8=1 - 16 q^{\frac12} + 128 q - 704 q^{\frac32} +\ldots\ .
\ee
Note that in (\ref{G4twists}) we have used the overall $\Z_2$ symmetry of the fixed point sectors to fix $\sigma(\infty)$ to have $\epsilon=0$, and that charge conservation fixes $\epsilon$ for $\sigma(1)$.
Let us expand (\ref{G4twists}) around $x=0$ to interpret the first few terms. 
To expand $q$ around $x=0$, we want to use the modular $S$ transform, since $x(-\textstyle\frac1\tau)=1-x(\tau)$. We then have
\be
\tilde q = q(-\textstyle\frac1\tau)= \frac{x^2}{256}+\frac{x^3}{256}+\frac{29 x^4}{8192}+\ldots
\ee
We have $Z_{\e_0,\e_1}(-\textstyle\frac1\tau)=Z_{\e_1,\e_0}(\tau)$.
Expanding around $x=0$, the contribution of the descendants of winding and momentum ground state $(m,n)$ is
\be
G(x)=\pm \frac{1}{x^{\frac18}\xb^{\frac18}}\frac{x^{h}\xb^{\hb}}{16^{h+\hb}}\left(1 +\frac{x^2}{128} + \frac{\xb^2}{128} +\ldots  \right) 
\ee
where
\be
h=\textstyle\frac12(\textstyle\frac mR+\textstyle\frac{nR}2)^2\ ,\qquad\hb=\textstyle\frac12(\textstyle\frac mR-\textstyle\frac{nR}2)^2\ .
\ee
We interpret this in the following way: $\sigma$ has dimension $(\textstyle\frac1{16},\textstyle\frac1{16})$, which leads to the overall factor in front. The contributions then come from winding and momentum states in the untwisted sector. The winding and momentum state $|p_L,p_R\rangle$ gives the leading contribution. The next terms come from the states $\alpha_{-1}\alpha_{-1}|p_L,p_R\rangle$ and $\tilde \alpha_{-1}\tilde \alpha_{-1}|p_L,p_R\rangle$ running in the intermediate channel. Those are indeed invariant under the orbifold, and thus survive in the untwisted sector. Note that there is no contribution of the state $ \alpha_{-1}\tilde \alpha_{-1}|p_L,p_R\rangle$, since that state has a vanishing 3-point function. This can be seen by computing the 3-point function on the cover: It is simply a 1-point function with one holomorphic and one antiholomorphic boson, which therefore vanishes. 

Crucially, note from (\ref{Zoneboson}) that the vacuum sector with $h=\bar h=0$ only appears in the OPE if $\epsilon_1=0$. (Note that we exchanged $\epsilon_0$ and $\epsilon_1$ due to the $S$ transformation.) This means that for $R$ irrational, if $\epsilon_1\neq0$, it is impossible to have a marginal operator in the OPE, since $h$ and $\bar h$ are not going to be half-integer. This means that $\sigma(x)$ and $\sigma(0)$ have to be in the same fixed point sector to get a marginal operator. 
The result for four bosons is the same, we simply take the product of four copies (\ref{G4twists}) with appropriate $\vec \epsilon$.

\subsection{Spin fields}\label{app:fermtwist}

We work in the NS sector in what follows. To compute correlation functions of the orbifold theory, it is convenient to bosonise the fermionic fields. Let us first consider the simpler example discussed in sections 2 and 5 of \cite{Dixon:1986qv} where we have a single complex fermion $\Psi$. We bosonise the fermion as
\be\label{bsnsn_1cf}
\Psi= e^{iH}\ ,\qquad\bar\Psi= e^{-iH}\ , \qquad  \Psi \bar\Psi = i\partial H\ ,
\ee
where $H$ is a real bosonic field. One can check that the OPEs between $\psi$ and $\bar\psi$ have the correct form.

With this definition, the bosonised spin fields in the twisted sector of the $\mathbb Z_2$ are of the form
\be\label{bsnstn}
\sigma_f^\pm=e^{\pm\frac i2H}\ ,
\ee
and similarly, the right-moving spin fields are given by
\be\label{bsnstn_r}
\tilde \sigma_f^\pm=e^{\pm\frac i2\tilde H}\ .
\ee
These indeed introduce the correct branch cuts as we have
\be\label{bsnsn_S_psi}
\Psi(z)\;\sigma_f^-(0)\sim\frac1{z^{\frac12}}\,\sigma_f^+(0)\ ,
\ee
and is anti-periodic when taken around the origin once, as expected. To get the above OPE, we used
\be\label{exp_ope}
\no{e^{ik_1\cdot H(z,\bar z)}}\no{e^{ik_2\cdot H(0,0)}}=z^{k_1\cdot k_2}\bar z^{k_1\cdot k_2}\no{e^{i(k_1+k_2)\cdot H(0,0)}[1+O(z,\bar z)]}\ ,
\ee
and set $\alpha^\prime=2$. The OPE between a boson and an exponential field is of the form
\be\label{OPExexpx}
\partial H(z)\no{e^{ikH(w)}} =-ik\,\frac{\no{e^{ikH(w)}}}{z-w}+\cdots\ .
\ee

For K3 we have two complex fermions that we bosonise as
\bea
&&\Psi^1=e^{iH^1}\ ,\qquad\bar\Psi^1=e^{-iH^1}\ ,\\
&&\Psi^2=e^{iH^2}\ ,\qquad\bar\Psi^2=e^{-iH^2}\ ,\nonumber
\eea
where $H^1$ and $H^2$ are real bosonic fields. The bosonised spin fields are then of the form
\be
\sigma_f^\pm=e^{\pm\frac i2(H^1+H^2)}\ ,\qquad\tilde \sigma_f^\pm=e^{\pm\frac i2(\tilde H^1+\tilde H^2)}\ ,
\ee
and have conformal dimension $\textstyle\frac14$.

The advantage of bosonisation is that all fermionic correlation functions, even those including spin fields, can be evaluated using free bosons. Concretely, the correlation function of the product of exponentials is of the form, see \cite[appendix 6.A.]{DiFrancesco:1997nk}:
\be\label{exps_ii}
\Big\langle\prod_{i=1}^n:e^{A_i}:\Big\rangle=e^{\sum_{i,j=1,\,i<j}^n\langle A_i\,A_j\rangle}\ .
\ee
For $A=ik_i\cdot H^i(z_i)$, this reads
\be\label{exps_iii}
\Big\langle\prod_{i=1}^n:e^{ik_i\cdot H^i(z_i)}:\Big\rangle=\prod_{i,j=1,\,i<j}^n(z_i-z_j)^{k_i\cdot k_j}\ ,
\ee
where we again have set $\alpha^\prime=2$. Using the usual contour deformation tricks, we evaluate correlation functions of fermionic descendants. In practice, we will only need this for fermion zero modes such as $\pbl{1,2}_0 \sigma^-_f$. For this we obtain
\bea\label{fermphi_ope}
&&\bar\Psi^1_0 \sigma^-_f(x)=\oint_{x^\prime=x}\frac{dx^\prime}{2\pi i}\,\frac1{x^{\prime\frac12}}\bar\Psi^1(x^\prime)\sigma^-_f(x)\\
&&\qquad\qquad
=\oint_{x^\prime=x}\frac{dx^\prime}{2\pi i}\,\frac1{x^{\prime\frac12}}e^{-iH^1(x^\prime)}\,e^{\frac i2(H^1+H^2)(x)}\nonumber\\
&&\qquad\qquad=\,:e^{\frac i2(-H^1+H^2)(x)}:\ ,\nonumber
\eea
and likewise for the other fermion zero modes. This shows that, as expected, the fermion zero modes act on the doublet $\sigma_f^\pm$.

In what follows it will be useful to compute the 4-point function of the spin fields that appear in the moduli. Charge conservation essentially fixes the only non-vanishing correlation function to be
\bea\label{4pf_OOOO_bf_vi}
&&\
\langle \sigma_f^{--}(z_1,\bar z_1)\;\;\sigma_f^{--}(z_2,\bar z_2)\;\;(\sigma_f^{--})^\dagger(z_3,\bar z_3)\;\;(\sigma_f^{--})^\dagger(z_4,\bar z_4)\rangle\\
&&\qquad\;=\frac{|z_1-z_2|\,|z_3-z_4|}{|z_1-z_3|\,|z_1-z_4|\,|z_2-z_3|\,|z_2-z_4|}\ ,\nonumber
\eea
where the daggers denote the complex conjugate operators. The mnemonic here is that the terms in the numerator appear whenever the vacuum is not exchanged between two fields due to charge conservation. This rule generalizes to the case where we have fermion zero modes acting on some of the spin fields. Moreover, we can treat the left- and right-movers separately. 

In our computations in the main text, we will only need two results for the right-movers specialized to $0,x,1,\infty$, namely:
\be\label{feranti1}
\langle\tilde\sigma^-_f| \tilde\sigma^-_f(1)(\tilde\sigma^-_f)^\dagger(x)|\tilde\sigma^-_f\rangle= \frac{1}{\xb^{\frac12}(1-\xb)^{\frac12}}\ ,
\ee
and
\be\label{feranti2}
\langle\tilde\sigma^-_f| (\tilde\sigma^-_f)^\dagger(1)\tilde\sigma^-_f(x)|\tilde\sigma^-_f\rangle = \frac{\xb^{\frac12}}{(1-\xb)^{\frac12}}\ .
\ee

\section{Ward identities}\label{app_ward}
We use the superconformal Ward identities introduced by \cite{Dixon:1989fj} (\emph{e.g.}, their section 3.2, equation (3.14), and the discussion below it) to compute correlation functions, see also \cite[section 3.2]{Dijkgraaf:1990dj}. The Ward identity for the supercurrents is of the form
\bea\label{ward_G}
&&\Big\langle\oint\frac{dw}{2\pi i}\,\xi(w)G(w)\;\phi_1(z_1)\;\cdots\;\phi_n(z_n)\Big\rangle=\\
&&\qquad\qquad\qquad\qquad\qquad\sum_{i=1}^n\xi(z_i)\Big\langle\phi_1(z_1)\;\cdots\;\big(G_{-\frac12}\cdot\phi_i(z_i)\big)\;\cdots\;\phi_n(z_n)\Big\rangle=0,\nonumber
\eea
where $G$ is the supercurrent, $\phi_i$ are primary fields, and
\be\label{ward_xi}
\xi(w)=aw+b,
\ee
with $a$ and $b$ two arbitrary complex numbers. The superscript(s) of $G$ are suppressed above: it is $G^\pm$, $G^{\alpha A}$, and $G^{\dot\alpha\alpha}$ associated with the $\mathcal N=2$, small $\mathcal N=4$, and large $\mathcal N=4$, respectively. The important point here is that the conformal dimension of $G$ is $\textstyle\frac32$, and so the correlation function on the left hand side of (\ref{ward_G}) scales as $\textstyle\frac1{w^3}$ in the limit $w\to\infty$. Thus, in order to eliminate contributions from infinity to the contour integral, the numerator has to grow at most as $w^2$ and so $\xi(w)$ is of the form (\ref{ward_xi}). Moreover, we can choose the values of the two free parameters $a$ and $b$ of the vector field $\xi(w)$ (\ref{ward_xi}) to fix the value of $\xi(w)$ at two points on the sphere; \emph{e.g.}, at $w=z_1$ and at $w=z_n$, such that \cite{Dijkgraaf:1990dj,deBoer:2008ss}:
\be\label{ward_ab}
\xi(w=z_1)=1,\quad\mathrm{and}\quad\xi(w=z_m)=0,\qquad1<m\le n.
\ee
This yields
\be\label{ward_ab_ii}
a=\frac{1}{z_1-z_m},\qquad b=\frac{-z_m}{z_1-z_m},\qquad\xi(w)=\frac{w-z_m}{z_1-z_m}.
\ee

The exactly marginal operators of the SCFTs discussed so far are schematically of the form $G_{-\frac12}\cdot\phi$, where $\phi$ is a chiral or anti-chiral primary with dimension $h=\textstyle\frac12$. We have again suppressed the superscripts of the supercurrents and the (anti)-chiral primaries corresponding to the symmetry group of the theory. For the correlation functions which contain one exactly marginal operator --- say, at $z_1$ --- the superconformal Ward identity (\ref{ward_G}) amounts to deforming the contour integral around the weight-$\textstyle\frac12$ (anti)-chiral primary at $z_1$ and circling around all the remaining positions, $z_i$, $i\in\{2,\cdots,n\}$ and infinity. The contribution from infinity has been taken into account by equation (\ref{ward_xi}). Then imposing the condition (\ref{ward_ab}) and using equation (\ref{ward_ab_ii}), the contour integral of the supercurrent can be written as \cite{Dixon:1989fj}
\be\label{ward_G_contour}
G_{-\frac12}\cdot\phi_i(z_i)=\oint_{w=z_i}\frac{dw}{2\pi i}\,\frac{w-z_m}{z_i-z_m}\;G(w)\phi_i(z_i).
\ee
This immediately yields $\xi(w=z_1)=1$. At $z=z_m$, since the OPE of $G$ with the primary operator $\phi_m$ has a simple pole $\textstyle\frac{1}{w-z_m}$, the pole is canceled out by the numerator of the integrand and so the contour integral vanishes, resulting in $\xi(w=z_m)=0$, as expected.

\small\baselineskip=.87\baselineskip
\let\bbb\bibitem\def\bibitem{\itemsep1.5pt\bbb}

\bibliographystyle{utphys}
\bibliography{refmain}

\end{document}